\documentclass[useAMS,usenatbib]{mn2e_mod}
\usepackage{natbib}
\usepackage{graphicx}
\usepackage{amsmath}
\usepackage{amssymb}
\usepackage{deluxetable}
\usepackage{longtable}

\newcommand{\Msol}{M$_{\odot}$}
\let\ACMmaketitle=\maketitle
\renewcommand{\maketitle}{\begingroup\let\footnote=\thanks \ACMmaketitle\endgroup}

\title [AGN feedback in E+A galaxy]{Evidence of ongoing AGN-driven feedback in a quiescent post starburst E$+$A galaxy \footnote{The data presented herein were obtained at the W.M. Keck Observatory, which is operated as a scientific partnership among the California Institute of Technology, the University of California and the National Aeronautics and Space Administration. The Observatory was made possible by the generous financial support of the W.M. Keck Foundation.}}

\author[Baron et al.]
{Dalya Baron$^{1}$\thanks{dalyabaron@mail.tau.ac.il},
Hagai Netzer$^{1}$,
Dovi Poznanski$^{1}$,
J. Xavier Prochaska$^{2}$,
\newauthor 
\& Natascha M. F{\"o}rster Schreiber$^{3}$\\
\\
$^{1}$School of Physics and Astronomy, Tel-Aviv University, Tel Aviv 69978, Israel.\\
$^{2}$Department of Astronomy and Astrophysics, UCO/Lick Observatory, University of California, 1156 High Street, Santa Cruz, CA 95064, USA.\\
$^{3}$Max-Planck-Institut f{\"u}r Extraterrestrische Physik, Giessenbachstrasse 1, D-85748 Garching, Germany.
}

\begin{document}

\maketitle

\label{firstpage}
\begin{abstract}

Post starburst E+A galaxies are thought to have experienced a significant starburst that was quenched abruptly. Their disturbed, bulge-dominated morphologies suggest that they are merger remnants. We present ESI/Keck observations of SDSS J132401.63+454620.6, a post starburst galaxy at redshift $z=0.125$, with a starburst that started 400 Myr ago, and other properties, like star formation rate (SFR) consistent with what is measured in ultra luminous infrared galaxies (ULRIGs). The galaxy shows both zero velocity narrow lines, and blueshifted broader Balmer and forbidden emission lines (FWHM=$1350\pm240\,\mathrm{km\,s^{-1}}$). The narrow component is consistent with LINER-like emission, and the broader component with Seyfert-like emission, both photoionized by an active galactic nucleus (AGN) whose properties we measure and model. The velocity dispersion of the broad component exceeds the escape velocity, and we estimate the mass outflow rate to be in the range 4--120 \Msol/yr. This is the first reported case of AGN-driven outflows, traced by ionized gas, in post starburst E+A galaxies. We show, by ways of a simple model, that the observed AGN-driven winds can consistently evolve a ULIRG into the observed galaxy. Our findings reinforce the evolutionary scenario where the more massive ULIRGs are quenched by negative AGN feedback, evolve first to post starburst galaxies, and later become typical red and dead ellipticals. 

\end{abstract}


\vspace{1cm}
\section{Introduction}\label{s:intro} 

E+A galaxies, also called $\mathrm{H\delta}$-strong, and K+A galaxies, show prominent Balmer absorption lines and a narrow stellar-age distribution, with a significant contribution from intermediate-age stars (typically A-type stars), indicating a recent starburst. Such systems show little or no contribution from younger stars (O and B type), with evidence for an abrupt termination of the star formation (SF) process \citep{dressler99, poggianti99, goto04, dressler04}.The estimated star formation rates (SFRs) during the starburst range from 50 to 300 $\mathrm{M_{\odot}/yr}$ \citep{poggianti00, kaviraj07}, and the mass fractions forming in the burst are very high, 30--80\% of the total stellar mass in the galaxy \citep{liu96, bressan01, norton01, yang04, kaviraj07}. 

Many E+A galaxies have bulge-dominated morphologies, with observed tidal features or close companions, suggestive of a late-stage merger \citep{canalizo00, yang04, goto04, cales11}. This evolutionary stage must be very short, due to the short life-time of A-type stars, making such systems rare and comprising only $\sim$3\% of the general galaxy population \citep{goto03, goto07, wild09, alatalo16a}, with an overall spatial distribution which follows that of the general galaxy population in the local universe \citep{blake04}. At intermediate redshifts ($0.3 < z < 1$) these galaxies appear to be four times more abundant in clusters, while in the local universe they have been mostly detected in the field \citep{zabludoff96, tran04, quintero04}.

Various studies suggest that post starburst galaxies provide an evolutionary link between the gas-rich star forming population ('blue cloud') and gas poor quiescent galaxies ('red sequence'; e.g., \citealt{yang04, yang06, kaviraj07, wild09, cales11, cales13, yesuf14, cales15, alatalo16a, alatalo16b, wild16}). In this scenario, a gas-rich major merger triggers a powerful starburst, which can be observed as an ultra luminous infrared galaxy (ULIRG), with SFRs ranging between 100 and 1000 $\mathrm{M_{\odot}/yr}$. The system then evolves into a post starburst galaxy (e.g., \citealt{kaviraj07, cales15}) which, in turn, evolves to a quiescent elliptical galaxy \citep{yang04, yang06, wild09, wild16}. \citet{yesuf14}, \citet{rowlands15}, and \citet{alatalo16b} studied the dust and molecular gas content in samples of E+A galaxies and generally found significantly more molecular gas and dust, compared to quiescent galaxies. This may indicate a transition between dusty ULIRGs and dustless red and dead galaxies. Note that in an E+A sample with Seyfert-type emission lines, \citet{yesuf17} find low gas mass fractions relative to normal star forming galaxies.

Numerical simulations show that gas rich major mergers are capable of triggering a significant starburst, and gas inflow onto the super massive black hole (SMBH) in the center of the galaxy. This can trigger an active galactic nucleus (AGN) that lasts as long as the gas supply from the galaxy is sufficient to maintain its activity \citep{mihos94, bekki01, springel05a, springel05b, hopkins06}. Various processes can quench the massive, ongoing, SF in such systems (negative feedback). Common sources of negative feedback are supernova- (SN) and AGN-driven winds, which can heat the gas via shocks, and photoionize it \citep{heckman90, dimatteo05, springel05a, ciotti10, feruglio10, cicone14}. Feedback can also affect the gas supply to the vicinity of the central SMBH.

There are several consequences to this evolutionary scenario. First, one would expect to observe post starburst galaxies with evidence of AGN activity, due to the time delay between the end of the starburst event and the termination of the accretion onto the BH. Indeed, \citet{goto06} identified 840 galaxies with both a post starburst signature and an AGN. These account for 4.2\% of all the galaxies in their volume-limited sample. \citet{wild10} and \citet{yesuf14} studied the growth of BHs in post starburst galaxies and found a rise in activity roughly 200-250 Myr after the onset of the starburst. Furthermore, a sub-class of AGN, known as post starburst quasars (PSQ), show both strong Balmer absorption lines, broad emission lines from the broad-line region (BLR), and a characteristic continuum emission typical of thin accretion disks (ADs). The first of these objects was discovered by \citet{brotherton99}. \citet{canalizo00} then identified a companion galaxy at the same redshift and concluded that the post starburst and the quasar activity are due to a recent interaction between the two (see also \citealt{brotherton02}). A sample of 29 PSQs images was studied by \citet{cales11} who found an equal number of spiral (13/29) and early-type (13/29) host galaxies. In a following work, \citet{cales13} studied the optical spectra of 38 such objects, and found BHs with masses of $\mathrm{M_{BH}\sim10^{8}\,M_{\odot}}$ that accrete at a few percent of the Eddington luminosity. They also found that PSQs hosted by spiral galaxies have older starburst ages and evidence of ongoing star formation, compared to their early-type counterparts. They therefore conclude that only the PSQs with early-type hosts are likely to result from major mergers. \citet{cales15} found that the more luminous PSQs (those hosted by ellipticals and early-types) display similar spectral properties to post starburst galaxies, but with older starburst ages. 

\citet{kaviraj07} performed the first comprehensive observational and theoretical study of the quenching process that terminates the starburst in E+A galaxies. They found a clear bimodal behaviour. For galaxies less massive than $\mathrm{M=10^{10}\,M_{\odot}}$, SN-driven feedback is the main quenching mechanism, while galaxies above this mass are mainly quenched by an AGN. They suggested, based on calculated star formation histories (SFHs), that the latter indeed were ULIRGs in their past. The less massive post starburst galaxies are expected to show signatures of SN-driven winds, and shocked gas signatures. More massive post starburst galaxies, in which the AGN is the main source of quenching, are expected to show signatures of photoionized gas and AGN-driven outflows. According to this scenario, an AGN which remains active after the termination of the burst is likely to show observational evidence for feedback, perhaps in a form of outflow. Such evidence is yet to be found in E+A galaxies.

There have been several attempts to look for shocks induced by SN- or AGN-driven winds in post starburst galaxies. \citet{alatalo16a} constructed a sample of $\sim$1\,000 E+A galaxies with emission line ratios that are consistent with shocks, and compared their properties to star forming, and AGN-dominated galaxies. Unfortunately, general shock models predict emission line ratios that can cover a very large range of emission line ratios (e.g., see figure 3 in \citealt{alatalo16a}). Thus, separating shock excitation from photoionization require a detailed study of the gas temperature and the general energetics of the shock.

There is very little, if any, clear observational evidence for AGN-driven outflow in local E+A galaxies. \citet{tremonti07} report on the detection of neutral gas outflows (using the MgII absorption line) of 500-2000 $\mathrm{km\,s^{-1}}$ in 10 out of 14 post starburst galaxies at z=0.6 (see also \citealt{alatalo16b}). Furthermore, \citet{tripp11} detect "warm-hot" plasma at a distance of $>$ 68 kpc from what is probably a post starburst galaxy, using ultraviolet absorption spectroscopy. To the best of our knowledge, no other cases of neutral or ionized outflow were reported so far in E+A galaxies. On the other hand, there is plenty of evidence for massive outflows in ULIRGs (e.g., \citealt{rupke02, martin05, martin06, soto12, soto12b}) but these are not connected directly with the post starburst phase of such systems. Obviously, there are thousands of AGNs that show some indication of outflow, such as the large SDSS type-II sample studied by \citet{mullaney13} and the more detailed study of 16 of those by \citet{harrison14}. However, there is no direct link between these outflows and the specific evolutionary phase of E+A galaxies.

SDSS J132401.63+454620.6 was found as an outlier galaxy by the anomaly detection algorithm of \citet{baron17}. This galaxy is one of the 400 spectroscopically weirdest galaxies in the 12th data release (DR12) of the Sloan Digital Sky Survey (SDSS). These include various extreme and rare phenomena such as galaxy-galaxy gravitational lenses, galaxies that host supernovae, double-peaked emission line galaxies, and more. 

In this paper we present our observations and analysis which show that SDSS J132401.63+454620.6 exhibits the first direct evidence of AGN-driven outflow in an E+A galaxy, traced by ionized gas emission. We estimate the mass outflow rate in the galaxy and show its possible connection with an earlier ULIRG phase and later evolution to quenched, early type galaxy. The paper is organized as follows. We describe the available SDSS observations and our high-resolution follow-up spectroscopy of SDSS J132401.63+454620.6 in section \ref{s:obs}. We measure various stellar and gas properties of the galaxy in section \ref{s:prop}, we then model the emission line gas in section \ref{s:comp_origin}. We characterise the outflow and discuss our results in the context of AGN feedback in the general galaxy population in section \ref{s:disc} and conclude in section \ref{s:concs}.

\section{Observations and Reductions}\label{s:obs}

\subsection{SDSS imaging and spectroscopy}
SDSS J132401.63+454620.6 was observed as part of the general SDSS survey \citep{york00}. The spectrum was obtained using a 3'' fiber with the SDSS spectrograph which covers a wavelength range of 3800\AA--9200\AA, with a resolving power going from 1500 at 3800\AA\, to 2500 at 9000\AA. The publicly available spectrum of the galaxy is combined from three 15m sub-exposures, and with signal to noise ratios (S/N) ranging from 4 to 25. We did not find spectral differences between the 3 sub-exposures. The image of the galaxy was taken in the SDSS broad band filters with a typical seeing of 1''--1.2''. 

\subsection{Keck/ESI spectroscopy}
We observed the galaxy using the Echelle Spectrograph and Imager (ESI; \citealt{sheinis02}) on the Keck II 10-m telescope on the night of 2016 June 2. Conditions were clear with seeing of 0.6''. The observations were carried out in echellette mode, which provides a wavelength coverage of 3900\AA\, to 10900\AA\, with a dispersion between 0.15 \AA/pix at 4200\AA\, and 0.39 \AA/pix at 9800\AA. The slit is 20'' in length and 0.75'' in width, providing a resolution of 44 $\mathrm{km\,s^{-1}}$. The pixel scale is 0.2''/pix. The Clear $S$ filter was used for the observations. Two exposures of 600 s each provided a combined spectrum with a median SNR of 12 per resolution element, ranging from 10 to 43. The spectra were reduced using the IDL-based ESIRedux\footnote{http://www2.keck.hawaii.edu/inst/esi/ESIRedux/index.html} pipeline \citep{prochaska03, bernstein15}.

\begin{figure*}
\includegraphics[width=0.95\textwidth]{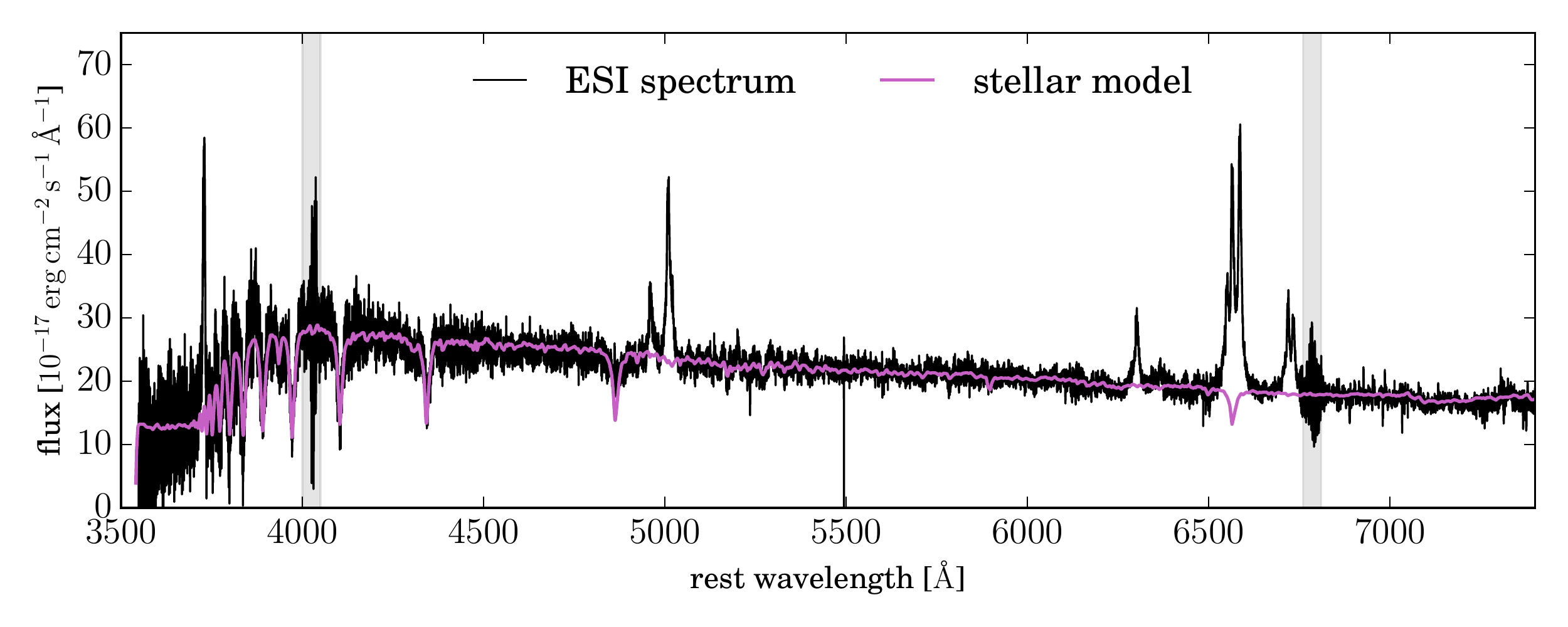}
\caption{The ESI spectrum of SDSS J132401.63+454620.6 (black), and the best population synthesis model (magenta) which we obtain using pPXF. Bad pixels are marked as grey bands.}\label{f:spec_and_pop}
\end{figure*}

Figure \ref{f:spec_and_pop} shows the combined ESI spectrum of J132401.63+454620.6. The spectrum is dominated by strong stellar features, typical of A-type stars, and a large number of strong emission lines typical of low and high ionization species. The strongest lines are: $\mathrm{[OIII]}$, $\mathrm{H\alpha}$, and $\mathrm{[NII]}$. The ESI spectrum is similar to the SDSS spectrum (i.e., we detect no variability during the 12 yr span between the two observations).

\section{Physical Properties}\label{s:prop}
In this section we present the physical properties of SDSS J132401.63+454620.6. We characterise the stellar properties in section \ref{s:st_prop} and the emission line properties in section \ref{s:emission_prop}.

\subsection{Spatial Distribution}\label{s:2d_spec}

\begin{figure}
\includegraphics[width=3.25in]{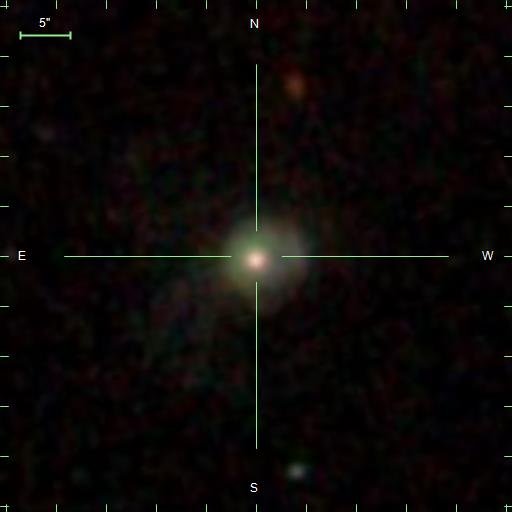}
\caption{The $gri$ color-composite image of SDSS J132401.63+454620.6.}\label{f:sdss_image}
\end{figure}

The SDSS color-composite image (figure \ref{f:sdss_image}) shows a bulge-dominated morphology and an additional feature that might be a tidal arm, suggestive of a late-stage merger. We fit a Sersic profile to the galaxy images in the $g$ and $r$ bands and find good fits with Sersic indices of 5.1 and 4.3 respectively. These values are consistent with the NYU value added catalog that provides best fitting profiles to the SDSS galaxies \citep{blanton05}, which give Sersic indices of 5.2 and 4.2 for $g$ and $r$ respectively. These indices are different from what is typically found in elliptical and spiral galaxies. We measure the half-light radius of the galaxy in the $g$ and $r$ bands to be 1.2'', corresponding to 2.7 kpc, similar to the seeing during the observation.

We examined the ESI 2D spectrum to investigate the spatial distribution of line and continuum features in the cross dispersion direction. 
We focus on the region containing the $\mathrm{[OIII]}$ line since it is strong and isolated from sky emission lines. 
We extract columns of pixels in the cross-dispersion direction and compared their width ($\sigma$), fitted by a Gaussian, to the seeing. 
We find a continuous decrease in the width as we move from continuum-dominated regions to emission line-dominated regions, with widths that change from 0.9'' (2 kpc) in continuum dominated regions to  0.6'' (1.4 kpc) in the line-dominated region. The latter is similar to the typical seeing during the observation, thus we conclude that the emission lines are not spatially resolved in our spectrum.

\subsection{Stellar Properties}\label{s:st_prop}

We start by fitting stellar population synthesis models to the spectrum of J132401.63+454620.6. We used the Penalized Pixel-Fitting stellar kinematics extraction code (pPXF; \citealt{cappellari12}), which is a public code for extracting the stellar kinematics and stellar population from absorption-line spectra of galaxies \citep{cappellari04}. The code uses the MILES library, which contains single stellar population (SSP) synthesis models and covers the full range of the optical spectrum with a resolution of full width at half-maximum (FWHM) of 2.3\AA\, \citep{vazdekis10}. We show the best fitting stellar model in figure \ref{f:spec_and_pop}. The model has a mass-weighted average age of 0.25 Gyr with a standard deviation of 0.11 Gyr. The code also fits the dust reddening of the starlight, assuming a \citet{calzetti00} extinction law, giving a colour excess of $\mathrm{E}(B-V)=0.3\,\mathrm{mag}$. This is consistent with star forming galaxies (where $\mathrm{E}(B-V)$ is in the range 0.3--0.6 mag) but is high compared to quiescent galaxies \citep{kauff03b}. The spectrum is dominated by A-type stars, and we find no significant contribution from O and B-type stars, since the youngest age contributing to the model is $\sim$80 Myr. This suggests little or no ongoing star formation in the galaxy. The narrow stellar age distribution and the lack of ongoing star formation indicate that this is a post starburst galaxy. The equivalent width of $\mathrm{H\delta}$ absorption is 7.9\AA, in the typical range for E+A galaxies (see e.g., \citealt{goto07, alatalo16a}). The stellar velocity dispersion that we measure is $130\,\mathrm{km\,s^{-1}}$, consistent with the measurement obtained from the SDSS spectrum \citep{bolton12, thomas13}.

The SDSS provides a stellar mass measurement based on the broad-band photometry \citep{maraston13} of the galaxy, but it may be biased by contributions from the strong emission lines. We therefore use the best stellar population synthesis model to measure the stellar mass, and find stellar mass of $\mathrm{log\,M/M_{\odot}=10.5}$, consistent with the SDSS measurement. In the JHU-MPA catalog \citep{kauff03b, b04, t04}, SDSS J132401.63+454620.6 has a median SFR of $25\,\mathrm{M_{\odot}/yr}$ (16th percentile of $11\,\mathrm{M_{\odot}/yr}$ and 84th percentile of $50\,\mathrm{M_{\odot}/yr}$), and its D4000 index is $1.19\pm0.01$. The SFR is slightly above the main sequence of star forming galaxies \citep{kauff03b}. We note that the SFR given by the JHU-MPA catalog is measured using the D4000 index. Therefore, the SFR is an average value over a period of roughly 100 Myrs. Since we do not find evidence of ongoing SF in the galaxy (no contribution of O- and B-type stars), this suggests vigurous SF in the past 100 Myrs (see section \ref{s:dust_mol}).

\subsection{Emission Line Properties}\label{s:emission_prop}

\subsubsection{Line profiles}\label{s:fitting}

\begin{figure*}
\includegraphics[width=0.8\textwidth]{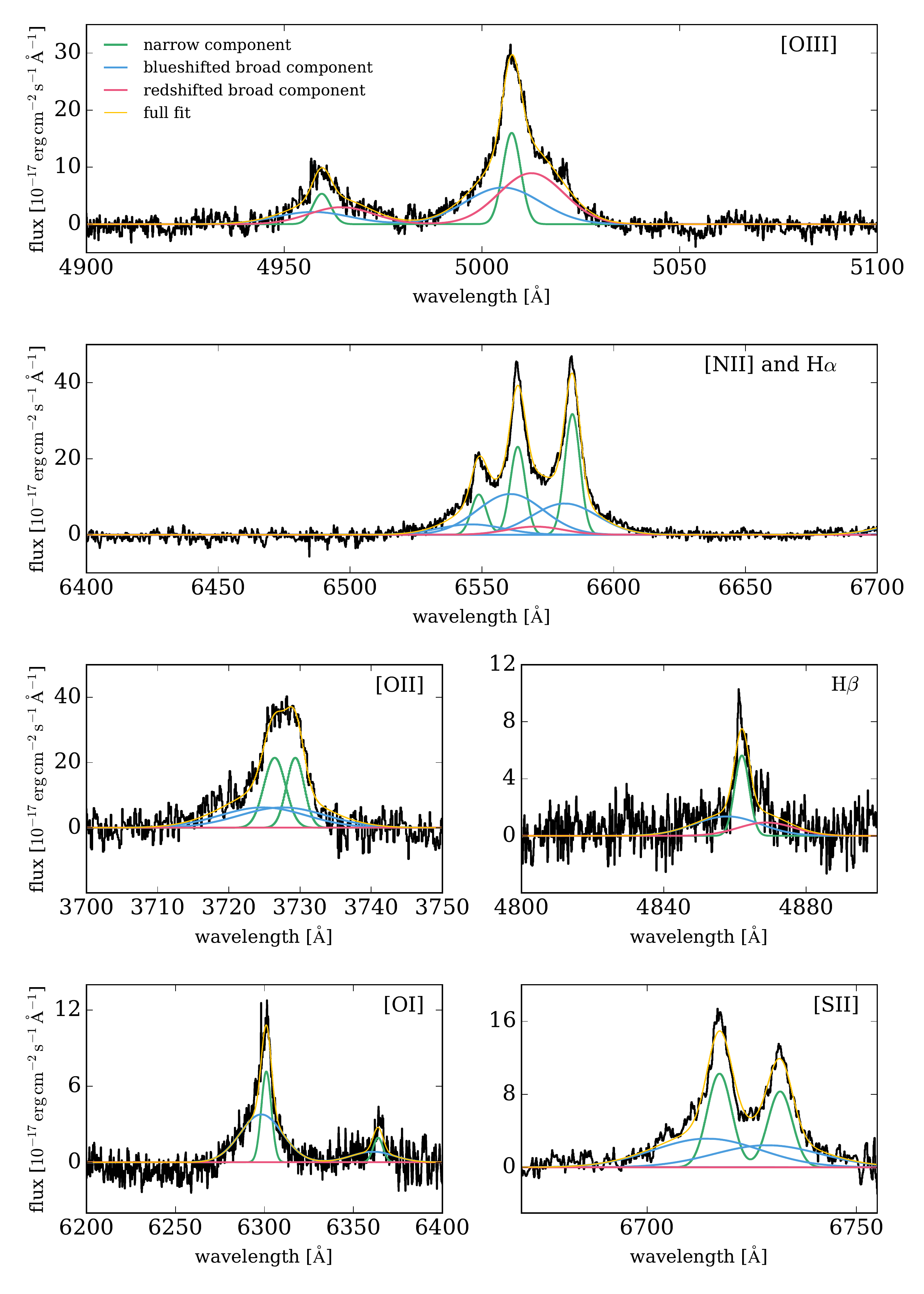}
\caption{The emission-line spectrum (black) of the gas in the galaxy as obtained by subtracting the best-fitting population synthesis model from the original spectrum. We show $\mathrm{[OIII]}$, $\mathrm{H\alpha}$ and $\mathrm{[NII]}$, $\mathrm{[OII]}$, $\mathrm{H\beta}$, $\mathrm{[OI]}$, and $\mathrm{[SII]}$ as indicated on each panel. We use three Gaussians to account for the narrow and broad components, and plot these in green (narrow lines), red (redshifted broad lines), and blue (blueshifted broad lines). The full fit is plotted in yellow. We refer the reader to section \ref{s:fitting} where we discuss the uncertainties in this fit.
}\label{f:emission_spec_fit}
\end{figure*}

We subtract the best-fitting population synthesis model and obtain the pure emission-line spectrum of the gas in the galaxy. 
Various parts of this spectrum are shown in figure \ref{f:emission_spec_fit}, where we show [OIII]~$\lambda \lambda$ 4959,5007\AA\, in the top panel, $\mathrm{H\alpha}$~$\lambda$ 6563\AA\, and [NII]~$\lambda \lambda$ 6548,6584\AA\, in the second panel, [OII]~$\lambda \lambda$ 3725,3727\AA\, and $\mathrm{H\beta}$~$\lambda$ 4861\AA\, in the third panel, and [OI]~$\lambda \lambda$ 6300,6363\AA\, and [SII]~$\lambda \lambda$ 6717,6731\AA\, in the bottom panel (hereafter $\mathrm{[OIII]}$, $\mathrm{H\alpha}$, $\mathrm{[NII]}$, $\mathrm{[OII]}$, $\mathrm{Hb}$, $\mathrm{[OI]}$, and $\mathrm{[SII]}$). 
Both the Balmer lines and the forbidden lines show narrow as well as broad components. 
The broad components of all lines, except $\mathrm{[OIII]}$ are clearly blueshifted with respect to the narrow core of the line. $\mathrm{[OIII]}$ is different and shows both blueshifted and redshifted broad components. 

Our de-blending procedure starts with the $\mathrm{[OIII]}$ line which has the highest S/N and is isolated from other emission lines. 
We fit three Gaussians to account for the emission - a narrow Gaussian, and two broad Gaussians, one redshifted and one blueshifted with respect to the narrow component. 
We tie the central wavelengths and widths of the Gaussians for [OIII]~$\lambda \lambda$4959,5007\AA\, and force the intensity ratio to the theoretical value of 3:1. We plot the best fit in the top panel of figure \ref{f:emission_spec_fit}, the narrow component is in green, the broad redshifted component is red, and the broad blueshifted component is blue. The full fit is yellow. 

We use the central wavelengths and widths obtained from the $\mathrm{[OIII]}$ line fit to fit the profiles of $\mathrm{[OII]}$, $\mathrm{H\beta}$, $\mathrm{[OI]}$, $\mathrm{H\alpha}$, $\mathrm{[NII]}$, and $\mathrm{[SII]}$. We find that all the narrow components have the same redshift and width. For the broad components, we force the redshifts and the widths to be consistent with the best fitting parameters obtained for $\mathrm{[OIII]}$, while letting the amplitudes of the Gaussians vary. We also tie the intensities of the two $\mathrm{[NII]}$ lines to match their theoretical ratio. We show the best fitting models in the other panels of figure \ref{f:emission_spec_fit}, using the same colour scheme as for $\mathrm{[OIII]}$.

The best fitting models show no broad redshifted emission in $\mathrm{[OII]}$, $\mathrm{[OI]}$, $\mathrm{[NII]}$, and $\mathrm{[SII]}$. For $\mathrm{H\alpha}$ we find a small contribution from a broad redshifted emission. The main uncertainty of this procedure arises from the $\mathrm{H\alpha}$-$\mathrm{[NII]}$ profile, where the different contributions from the redshifted and blueshifted broad emission are blended into a single profile. However, although weaker, $\mathrm{[OII]}$, $\mathrm{[OI]}$, and $\mathrm{[SII]}$ are isolated and their ionization potentials are closer to that of $\mathrm{[NII]}$. Thus we expect the broad $\mathrm{[NII]}$ to resemble $\mathrm{[OII]}$, $\mathrm{[OI]}$, and $\mathrm{[SII]}$ profiles more than the $\mathrm{[OIII]}$ profile. 

We examine the effect of forcing a non-zero redshifted broad Gaussian in the fitting of the $\mathrm{[OII]}$, $\mathrm{[OI]}$, and $\mathrm{[SII]}$ profiles. We find a redshifted flux excess that exceeds the relevant line profiles, thus only a zero amplitude can be consistent with the line profiles, and it is less likely that there is $\mathrm{[NII]}$ redshifted broad contribution to the $\mathrm{H\alpha}$-$\mathrm{[NII]}$ profile. On the other hand, one can force a zero amplitude for the redshifted broad $\mathrm{H\alpha}$ component and still have a good fit. In this case the amplitudes of the blueshifted broad $\mathrm{H\alpha}$ and $\mathrm{[NII]}$ lines increase to account for the missing broad emission, and the amplitude ratio of $\mathrm{[NII]}$ and $\mathrm{H\alpha}$ remains roughly the same. It is possible that the redshifted broad $\mathrm{[OIII]}$ component is produced by gas whose level of ionization is too large to produce detectable $\mathrm{[NII]}$, $\mathrm{[SII]}$, $\mathrm{[OII]}$ and $\mathrm{[OI]}$ lines. However, it is impossible to emit $\mathrm{[OIII]}$ lines without some Balmer line emission. We therefore favour the best-fitting model in which the redshifted broad $\mathrm{H\alpha}$ emission contributes to the total line profile.

We also examined models with only two Gaussians to account for the narrow and broad emission in all the lines. In this case, we do not tie the central wavelengths of the broad Gaussians to reside at the same redshift, and a good fit can be achieved only if the broad $\mathrm{[OIII]}$ emission is redshifted by $\mathrm{310\,km\,s^{-1}}$ compared to the central wavelength of the broad forbidden (and Balmer) lines. This model results in a $\mathrm{[OIII]/H\beta}$ ratio of 40, much higher than what is observed in luminous AGN. The two Gaussians model has no effect on the isolated forbidden lines ($\mathrm{[OII]}$, $\mathrm{[OI]}$, $\mathrm{[SII]}$) and results in roughly the same $\mathrm{[NII]}$/$\mathrm{H\alpha}$ ratio as the three Gaussian case. In the following sections we measure different parameters of the gas and note the difference, if it exists, between the results obtained with the two and three Gaussian models.

We also examine the possibility that some of the redshifted emission that we see in $\mathrm{[OIII]}$ is due to $\mathrm{FeII}$ emission around 5018 \AA, as observed in many type-I AGN (e.g., \citealt{netzer07}). We use a theoretical $\mathrm{FeII}$ template described in \citet{netzer07} which is based on the \citet{boroson92} template and convolve it to fit the spectral resolution of the ESI spectrum. We restrict ourselves to $\mathrm{FeII}$ emission in the wavelength range 4900 - 5400\AA, and rescale the relative amplitude of the template to match the features in our spectrum. We find that $\mathrm{FeII}$ cannot account for any of the redshifted $\mathrm{[OIII]}$ emission while still being consistent with other strong $\mathrm{FeII}$ lines (such as $\mathrm{FeII\, 5160}$\AA), which are very weak in our spectrum.

Table \ref{t:meas} gives line fluxes, reddening corrected line luminosities, and derived fit parameters for all measured emission lines. The uncertainties of the widths and central wavelengths are obtained from the $\chi^{2}$ minimisation process, and are tied to each other as discussed above. We propagate these uncertainties and the uncertainties on the dust reddening (see next section) to calculate the uncertainties in line intensity and dust corrected luminosity.

\begin{table}
	\centering 
	\tiny
	\tablewidth{0.8\linewidth} 
\begin{tabular}{|p{2.8cm}||c|r|}
 Emission & Intensity & Luminosity \\
 Line  &  [$10^{-16}\,\mathrm{erg\,s^{-1}\,cm^{-2}}$]    &  [$\mathrm{erg\,s^{-1}}$]\\
 \hline
 \hline
 \multicolumn{3}{|l|}{Narrow lines: $\mathrm{\sigma=136 \pm 20\,km\,s^{-1}}$, $\mathrm{WL([OIII]\lambda5007)=5007.6\pm 0.2}$\AA} \\
 \hline
$\mathrm{H\beta}$				& $2.75 \pm 0.58$   & $1.27 \pm 0.27 \times 10^{41}$ \\
$\mathrm{[OIII]\lambda 4959}$   & $2.94 \pm 0.38$   & $1.27 \pm 0.17 \times 10^{41}$ \\
$\mathrm{[OIII]\lambda 5007}$   & $8.9 \pm 1.2$   & $3.75 \pm 0.50 \times 10^{41}$ \\
$\mathrm{[OI]\lambda 6300}$     & $4.93 \pm 0.77$   & $1.11 \pm 0.17 \times 10^{41}$ \\
$\mathrm{[NII]\lambda 6548}$    & $7.2 \pm 1.0$   & $1.53 \pm 0.22 \times 10^{41}$ \\
$\mathrm{H\alpha}$              & $17.1 \pm 2.2$   & $3.63 \pm 0.48 \times 10^{41}$ \\
$\mathrm{[NII]\lambda 6584}$    & $20.8 \pm 2.5$    & $4.58 \pm 0.54 \times 10^{41}$ \\
$\mathrm{[SII]\lambda 6717}$    & $8.92 \pm 0.46$   & $1.83 \pm 0.10 \times 10^{41}$ \\
$\mathrm{[SII]\lambda 6731}$    & $7.54 \pm 0.52$   & $1.54 \pm 0.12 \times 10^{41}$ \\
								&					&  \\
 \hline
 \hline
 \multicolumn{3}{|l|}{Blueshifted broad lines: $\mathrm{\sigma=572\pm 100 \,km\,s^{-1}}$, $\mathrm{WL([OIII]\lambda5007)=5005.4 \pm 1.2}$\AA} \\
 \hline
$\mathrm{H\beta}$				& $1.49 \pm 0.94$   & $4.2 \pm 2.7 \times 10^{42}$ \\
$\mathrm{[OIII]\lambda 4959}$   & $4.8 \pm 2.3$     & $11.4 \pm 5.5 \times 10^{42}$ \\
$\mathrm{[OIII]\lambda 5007}$   & $15.1 \pm 3.5$     & $32.3 \pm 7.7 \times 10^{42}$ \\
$\mathrm{[OI]\lambda 6300}$     & $8.8 \pm 2.7$     & $3.5 \pm 1.1 \times 10^{42}$ \\
$\mathrm{[NII]\lambda 6548}$    & $9.2 \pm 2.1$     & $3.11 \pm 0.71 \times 10^{42}$ \\
$\mathrm{H\alpha}$              & $35.1 \pm 5.8$    & $11.8 \pm 2.0 \times 10^{42}$ \\
$\mathrm{[NII]\lambda 6584}$    & $27.9 \pm 6.3$    & $9.3 \pm 2.1 \times 10^{42}$ \\
$\mathrm{[SII]\lambda 6717}$    & $9.8 \pm 4.2$     & $3.0 \pm 1.3 \times 10^{42}$ \\
$\mathrm{[SII]\lambda 6731}$    & $7.5 \pm 4.2$     & $2.3 \pm 1.2 \times 10^{42}$ \\
								&					&  \\
\hline
\hline
\multicolumn{3}{|l|}{Redshifted broad lines: $\mathrm{\sigma=492 \pm 90 \,km\,s^{-1}}$, $\mathrm{WL([OIII]\lambda5007)=5012.6 \pm 1.1}$\AA} \\
\hline 
$\mathrm{H\beta}$ 				& $1.26 \pm 0.75$   & $0.21 \pm 0.13 \times 10^{41}$ \\
$\mathrm{[OIII]\lambda 4959}$   & $5.8 \pm 2.1$   & $0.97 \pm 0.34 \times 10^{41}$ \\
$\mathrm{[OIII]\lambda 5007}$   & $18.1 \pm 6.3$   & $2.9 \pm 1.0 \times 10^{41}$ \\
$\mathrm{H\alpha}$              & $5.6 \pm 2.9$   & $0.62 \pm 0.32 \times 10^{41}$ \\
 								&					  &  \\
 \hline
\end{tabular}
\caption{Best fitting model parameters for the three Gaussian fit. The best fitting intensities are normalized with respect to the median value of the entire spectrum. The central wavelengths and widths of the Gaussians are tied together for a given velocity component. The luminosities are corrected for reddening by foreground ISM-type dust assuming $\mathrm{H\alpha/H\beta}$=2.85.}
\label{t:meas}
\end{table}

\subsubsection{Emission line gas properties}\label{s:gas}

Using the best fitting model, we derived the kinematic properties of the emitting gas. All the properties are given in table \ref{t:meas} with their uncertainties. The FWHM velocity that we measure for the narrow component is $320\,\mathrm{km\,s^{-1}}$, and the blueshifted and redshifted broad components have widths of $1300\,\mathrm{km\,s^{-1}}$ and $1200\,\mathrm{km\,s^{-1}}$ respectively. The peak of the blueshifted broad emission is shifted by $130\,\mathrm{km\,s^{-1}}$ with respect to the narrow component, and the redshifted is shifted by $300\,\mathrm{km\,s^{-1}}$. The narrow gas component is blueshifted by $50\,\mathrm{km\,s^{-1}}$ from the systemic redshift of the galaxy which we tie to the stellar Balmer absorption lines. 

One can compare the velocity dispersion that we measure in the broad lines to the escape velocity from the galaxy. To achieve that, we use the SDSS imaging of the galaxy and fit a Sersic profile to the light profile in $r$-band. We find a good fit with a Sersic index of 4.3. We assume a constant mass-to-light ratio as a function of distance from the center of the galaxy, which is justified by the dominance of A-type stars in the SDSS and ESI spectra. We examine two extreme cases, one assumes that the total mass is the stellar mass (no gas), and the second assumes gas mass that equals the stellar mass, resulting in a total mass of twice stellar mass. \citet{rowlands15} study the ISM in 11 post starburst galaxies and measure gas to stellar mass ratios which are consistent with normal galaxies. Thus, the real total mass must lie within this range. We then measure the line-of-sight FWHM of the escape velocity for these cases (see \citealt{agnello14}). The maximal line-of-sight FWHM of the escape velocity that we derive are $640\,\mathrm{km\,s^{-1}}$ and $970\,\mathrm{km\,s^{-1}}$ for the stellar mass and twice the stellar mass respectively. The redshifted and blueshifted broad components show a FWHM of $1300\,\mathrm{km\,s^{-1}}$ and $1200\,\mathrm{km\,s^{-1}}$ respectively, which exceed both values for the total mass.

\begin{figure*}
\includegraphics[width=0.95\textwidth]{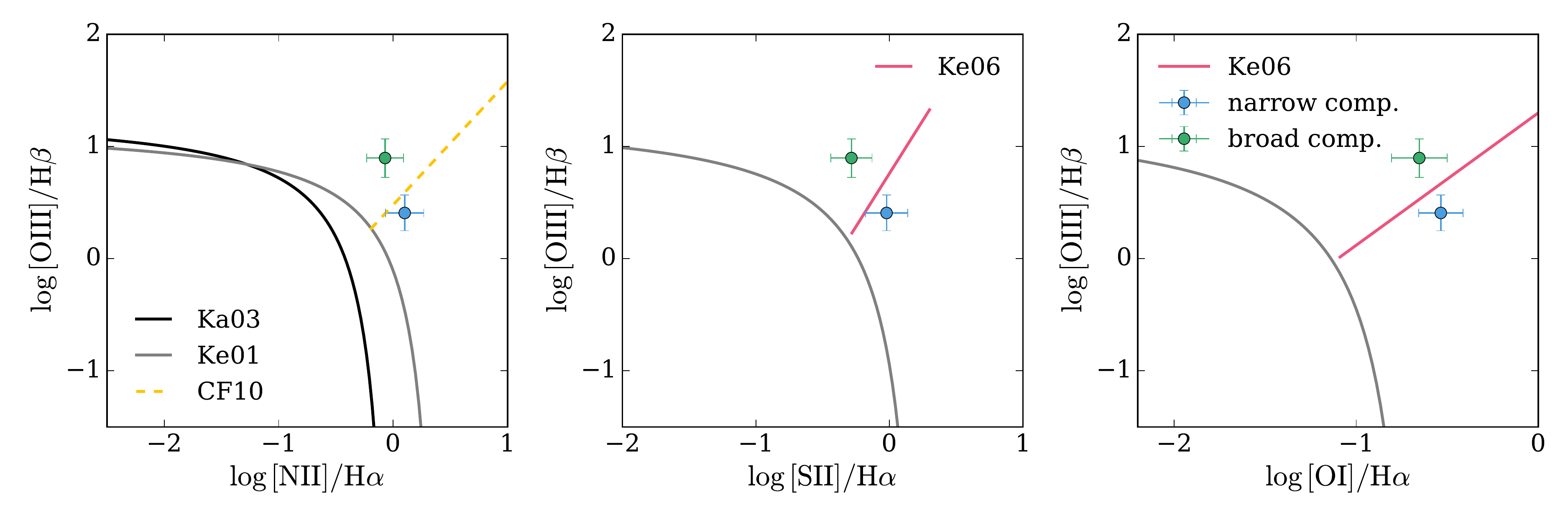}
\caption{Classification diagrams for the narrow (blue) and broad (green) emission lines. The left panel shows the emission line ratios $\mathrm{log\,[OIII]/H\beta}$ versus $\mathrm{log\,[NII]/H\alpha}$, where we mark the extreme starburst line by Ke01 with black, the composite line by Ka03 with grey, and the LINER-Seyfert separation by CF10 with yellow. The middle and right panels show the emission line ratios $\mathrm{log\,[OIII]/H\beta}$ versus $\mathrm{log\,[SII]/H\alpha}$ and $\mathrm{log\,[OI]/H\alpha}$ respectively. We mark the separation between LINER and Seyfert line (Ke06) by a pink line. In all diagrams, the narrow component is consistent with LINER and the broad component is consistent with Seyfert-like emission. 
}\label{f:bpt_components}
\end{figure*}

Next we examine the physical properties in the regions emitting the various line components, treating them as independent of each other. 
In figure \ref{f:bpt_components} we show the narrow (blue) and the broad blueshifted (green) emission components on several, standard, line-diagnostic diagrams \citep{baldwin81, veilleux87}. The left panel shows the emission line ratios $\mathrm{log\,[OIII]/H\beta}$ versus $\mathrm{log\,[NII]/H\alpha}$. We consider two separating criteria. The first is from \citet{kewley01} who used a combination of population synthesis models and photoionization models to produce a theoretical upper limit that separates starbursts and AGN-dominated galaxies (Ke01; black line). \citet{kauff03a} modified this limit by including composite galaxies which contain significant contribution from both star formation and AGN (Ka03; grey line). We use the LINER-Seyfert separating line from \citet[CF10]{cidfernandes10}. The diagram shows that both the narrow and the broad components are AGN-dominated. The middle panel and the right panel show the emission line ratios $\mathrm{log\,[OIII]/H\beta}$ versus $\mathrm{log\,[SII]/H\alpha}$ and $\mathrm{log\,[OI]/H\alpha}$ respectively. We also mark the \citet{kewley06} separating line between LINER and Seyfert galaxies (Ke06; pink line). In all diagrams, the narrow emission lines are consistent with LINER-like emission, and the broad components are consistent with Seyfert-like emission. We further note that when using two Gaussians instead of three Gaussians in the line fitting, we get a broad component with higher $\mathrm{[OIII]/H\beta}$ ratio, thus a line ratio which is further away from the LINER-Seyfert separation line in the diagram. The narrow component shows slightly smaller $\mathrm{[NII]/H\alpha}$, $\mathrm{[SII]/H\alpha}$ and $\mathrm{[OI]/H\alpha}$ ratios, though is still classified as a LINER. 

Next we attempt to derive the temperature and density of the emission line gas, and the reddening towards the two (broad and narrow) emitting regions.
The measured $\mathrm{H\alpha}$/$\mathrm{H\beta}$ ratio for the LINER component is 6.2, and for the Seyfert component is 23.4. Assuming an intrinsic line ratio of 2.85 and the \citet{cardelli89} extinction curve, we find reddening of $\mathrm{E}(B-V)=0.68\,\mathrm{mag}$ and $\mathrm{E}(B-V)=1.82\,\mathrm{mag}$ for the LINER and Seyfert components, respectively (we find differences of at most 10\% when using SMC or LMC extinction curves e.g., \citealt{baron16}). These values are higher compared to the reddening found in our stellar synthesis model for the host galaxy, $\mathrm{E}(B-V)=0.3\,\mathrm{mag}$. The latter was obtained with the \citet{calzetti00} reddening law which is more appropriate for the stellar population.

We use the narrow and broad [SII]~$\lambda$6716\AA/[SII]~$\lambda$6731\AA\, line ratios to calculate the electron density in these gas components \citep{osterbrock06}, and find $200\pm50\,\mathrm{cm^{-3}}$ for the LINER component and $100\pm40\,\mathrm{cm^{-3}}$ for the Seyfert component. 

Measuring the electron temperature of the emission line gas is more problematic, since we do not detect [OIII]~$\lambda$4363\AA\, and [NII]~$\lambda$5755\AA, and hence can only place an upper limit on the electron temperature \citep{osterbrock06}. Moreover, this can be done only for the LINER component, since the S/N of the Seyfert component is too low. We first fit single Gaussians to the undetected [OIII]~$\lambda$4363\AA\, and [NII]~$\lambda$5755\AA\, lines and correct the measured intensities for dust reddening. Since these lines are undetected, the dust-corrected intensities serve as upper limits for [OIII]~$\lambda$4363\AA\, and [NII]~$\lambda$5755\AA. To measure the temperature that corresponds to this limit, we use the intensity ratios $\mathrm{[OIII]\, (4959+5007)/4363}$ and $\mathrm{[NII]\, (6548+6584)/5755}$. We find upper limits of 25\,000 K and 18\,000 K from $\mathrm{[OIII]}$ and $\mathrm{[NII]}$ respectively. The difference between the upper limits is mainly due to the difference in the strengths of [OIII]~$\lambda$5007\AA\, and [NII]~$\lambda$6584\AA. An upper limit of 18\,000 K is consistent with shocked gas with velocities of 200-600 $\mathrm{km\,s^{-1}}$ \citep{allen08}. We perform a detailed comparison with shock models in section \ref{s:comp_origin}.

\section{Modelling the emission line gas}\label{s:comp_origin}

We show in section \ref{s:gas} and in figure \ref{f:bpt_components} that the emission line spectrum of SDSS J132401.63+454620.6 exhibits two separate components, one which is consistent with the spectrum of LINERs and the other with the spectrum of Seyfert 2 galaxies. The main difference between the two is in the ratio of $\mathrm{L([OIII])/L(H\beta)}$, which is a factor of about 3 larger in the broad, blueshifted ($\sim 10$) component compared with the narrow component ($\sim 3$). The differences in the three other ratios, $\mathrm{[NII]/H\alpha}$, $\mathrm{[OI]/H\alpha}$, and $\mathrm{[SII]/H\alpha}$, are smaller, but all three are clearly stronger in the narrow cores of the lines. While it is not unusual for an active galaxy to show contribution from both HII regions and AGN (e.g., \citealt{kewley06}), galaxies with both LINER and Seyfert components were not perviously reported, to the best of our knowledge. 

If all the broad and narrow emission line components are due to ionization by an AGN radiation field, one must come up with a specific geometry that will produce both LINER and Seyfert emission. An alternative scenario involves contribution from both AGN and shocked gas, or contributions from shocked gas at different velocities \citep{dopita97, nagar05}. An additional process that can produce LINER-like emission is photoionization by evolved post asymptotic giant branch stars (post AGB stars) during a short but very hot and energetic phase \citep{stasinka08, annibali10, cidfernandes11, yan12, singh13}.

In this section we examine the ionization source of the emission lines. We investigate shock-induced emission lines (section \ref{s:shocks}), photoionization by post AGB stars (section \ref{s:p_agb}), and photoionization by an AGN (section \ref{s:agn_ion}). For simplicity, in the following discussion we refer to the broad, blueshifted components of all emission lines as the "Seyfert type emission", because of its location on the line-diagnostic diagrams. However, this is rather different from most Seyfert 2 galaxies because the lines are very broad and shifted with respect to the systemic velocity and hence may not represent the conditions in a "typical" Seyfert-like narrow line region (NLR). We explain this issue in detail in section \ref{s:agn_ion}. 

\subsection{Shock excitation}\label{s:shocks}
Winds from supernovae or AGN-driven outflows can produce shocked gas by injection of mechanical energy into the ISM. The emission-line spectrum of shocked gas is different from gas that is photoionized by massive stars, but share similarities with the spectrum of gas that is photoionized by an AGN. Both of these manifest themselves as emission line ratios that are consistent with LINER- or Seyfert-like spectra. Indeed, various studies use emission-line ratios as shocked gas signatures (e.g., \citealt{alatalo16a, alatalo16b, soto12, soto12b}). Such line ratios can cover a very large region in the line diagnostic diagrams considered in section \ref{f:bpt_components} (see e.g., \citealt{alatalo16a}) to the extent that line ratios by themselves cannot be used to uniquely identify the shock signature (see however \citealt{soto12b}). 

Here we use general energetics considerations based on the fact that the relative efficiency of collisional ionization in shocks and photoionization by the AGN radiation field is proportional to $v^2/c^2$ (see e.g., \citealt{laor98}). In the specific case considered here, the two processes involve mass flow rates that differ by about three orders of magnitude. That is, the AGN accretion rate that is necessary to produce the observed emission lines is three orders of magnitudes lower than the mass outflow rate of a wind which is required to produce the same emission-line luminosities via shocks.

To illustrate the limitations of shock excited models, we investigated a specific example  that applies to our object. We use a grid of fast shock models from MAPPINGS III \citep{dopita95, dopita96, dopita05, allen08}. \citet{allen08} provide a library of fully radiative shock models with shock velocities in the range $v_{s} = 100 - 1000\, \mathrm{km\,s^{-1}}$, magnetic parameters of $\mathrm{B/n^{1/2}=10^{-4}-10\,\mu G\, cm^{3/2}}$, and a set of five different atomic abundances, for a pre-shock density of $0.1\,\mathrm{cm^{-3}}$. For solar abundance, they provide shock models with pre-shock densities of $0.01 - 1000\,\mathrm{cm^{-3}}$. For each model, they provide all details about the radiative shock and its photoionized precursor, including their temperatures and luminosities. The library consists of emission line ratios compared to $\mathrm{H\beta}$. We use all the models that are available in the library: shock alone, precursor alone, and a combination of shock and precursor, for the entire range of parameters: density, temperature, magnetic parameter, and abundance.

We first compared four measured emission line ratios $\mathrm{[OIII]/H\beta}$, $\mathrm{[NII]/H\alpha}$, $\mathrm{[SII]/H\alpha}$, and $\mathrm{[OI]/H\alpha}$ to the emission line ratios predicted by the shock models, keeping models which agree with the observations in all four line ratios, within 0.15 dex. These emission line ratios are not sensitive to the dust reddening, and the comparison does not depend on the assumed $\mathrm{H\alpha/H\beta}$ line ratio. We find multiple shock models with emission line ratios that are consistent with the measured ratios. We note that in the two Gaussian case, where the Seyfert component has an extremely high $\mathrm{[OIII]/H\beta}$ ratio, no model reaches the measured ratio to within 0.5 dex.

For the LINER component, we find that shock models with velocities ranging between 300 -- 450 $\mathrm{km\,s^{-1}}$, a density of 1 $\mathrm{cm^{-3}}$, magnetic parameters of 0.5--6.32 $\mathrm{\mu G\, cm^{3/2}}$, and solar and twice solar abundances, predict line ratios that are consistent with the observations. For the Seyfert component, the shock models with velocities of 425 -- 525 $\mathrm{km\,s^{-1}}$, densities of 0.01--1 $\mathrm{cm^{-3}}$, magnetic parameters of 0.5--10 $\mathrm{\mu G\, cm^{3/2}}$, and solar and twice solar compositions match the observed line ratios very well. The electron temperatures of these models, 10\,000--14\,000 K, are consistent with the upper limit, 18\,000 K, that we derive in section \ref{s:gas} for the narrow line component. Taking into account the measured reddening, the same shock models also produce the required relative intensities of [NI]~$\lambda$5200\AA, [OII]~$\lambda \lambda$3726,3728\AA\, and [OI]~$\lambda$6364\AA.

The MAPPINGS III library also provides the predicted $\mathrm{H\beta}$ luminosity per unit area for the different shock models. We use the 10 models that are most consistent with the measured LINER and Seyfert line ratios and extract the predicted $\mathrm{H\beta}$ luminosity per unit area. The predicted luminosity ranges between $\mathrm{log\,L(H\beta)=-4\,\,[erg\,cm^{-2}s^{-1}]}$ and $\mathrm{log\,L(H\beta)=-5.3\,\, [erg\,cm^{-2}s^{-1}]}$. The emitting area that is required to produce the observed $\mathrm{H\beta}$ luminosity ranges from 30 to 1000 $\mathrm{\mathrm{kpc^{2}}}$. As noted earlier, the emission lines are not resolved in the 2D spectrum of the galaxy, leading to a maximum angular size of 0.6'', which is 1.4 kpc, or a maximum emitting area of about 2 $\mathrm{\mathrm{kpc^{2}}}$. This size is several orders of magnitude smaller than the size predicted by the shock models. 

We can further examine the mechanical energy produced by a shock that can produce the measured line luminosity. The mechanical energy produced by the shock is $1/2\,\mathrm{m_{sh}v_{sh}^{2}}$, where $\mathrm{m_{sh}}$ is the mass of the flowing material and $\mathrm{v_{sh}}$ is the shock velocity. We can assign an efficiency factor to this process, $\mathrm{\eta_{sh}=1/2\,(v/c)^2}$. In our case, the highest possible $\mathrm{\eta_{sh}}$, corresponding to the highest shock velocity that produces emission line ratios that are consistent with observations, is $1.5 \times 10^{-6}$. The energy that is associated with photoionisation by AGN due to mass accretion onto the BH is $\mathrm{\eta_{acc}m_{acc}c^2}$, where $\mathrm{\eta_{acc}}$ is the mass-to-energy conversion efficiency of the accretion process, and we take it to be 0.1. Therefore, the energy ratio produced by these two processes is $\mathrm{(\eta_{acc}m_{acc}) / (\eta_{sh}m_{sh})}$. 
We measure in section \ref{s:agn_props} the mass accretion rate onto the BH, $\dot{m}_{acc} \approx 0.17\,$ \Msol/yr. Given the assumed efficiencies of the two processes, a mass flow rate of over $10\,000\,$ \Msol/yr is necessary to produce the measured line luminosities via shocks. This number is several orders of magnitude larger than any mass outflow rate observed in similar systems and about a factor of 100 larger than what is deduced below (section \ref{s:agn_quench}) for SDSS J132401.63+454620.6. We therefore conclude that shock models cannot produce the emission lines observed in this source. 

\subsection{post-AGB ionization}\label{s:p_agb}

\citet{singh13} use integral field spectroscopic data from the CALIFA survey and study the observed radial surface brightness profiles of galaxies that show LINER-like emission, which they compare to what is expected from illumination by an AGN. They find that the radial emission-line surface brightness profiles are inconsistent with ionization by a central point-source and hence cannot be due to an AGN alone. They propose post-AGB stars as the ionizing source in many objects classified as LINERs. In a related work, \citet{cidfernandes11} study such cases and provide an estimated upper limit to EW($\mathrm{H\alpha}$) of about 3\AA\, for post-AGB systems. This is considerably smaller than the EW obtained for SDSS J132401.63+454620.6 (about 8\AA). Moreover, post-AGB stars are ubiquitous in galaxies with stellar populations older than $\sim 1\,\mathrm{Gyr}$. In SDSS J132401.63+454620.6 the spectrum is dominated by A-stars and its stellar age spans the range 0.02-0.1 Gyr with no detected contribution from older stars. Thus it is unlikely that the LINER-like emission is due to post-AGB stars. 

\subsection{AGN ionization}\label{s:agn_ion}

We show in sections \ref{s:shocks} and \ref{s:p_agb} that the narrow (LINER) and broad (Seyfert) emission lines in SDSS J132401.63+454620.6 are inconsistent with ionization and excitation by shocks or post AGB stars. We therefore suggest that the lines arise due to the radiation field of a central active BH. Below we present the basic AGN properties derived from the observations and suggest a model that explains the emission line properties.

\subsubsection{AGN properties}\label{s:agn_props}

The dust-corrected [OIII]~$\lambda$5007\AA\, and [OI]~$\lambda$6300\AA\, luminosities for the LINER and the Seyfert components are given in table \ref{t:meas}. The bolometric luminosity of the AGN is measured as follows \citep{netzer09}:
\begin{equation}\label{eq:1}
	{\mathrm{log}\,L_{\mathrm{bol}} = 3.8 + 0.25\mathrm{log}L(\mathrm{[OIII]}) + 0.75\mathrm{log}L(\mathrm{[OI]})}
\end{equation}
We find $L_{bol}=10^{45}\,\mathrm{erg\,s^{-1}}$ for the narrow emission line component. The luminosity deduced from the broad component, using the same relationship, is an order of magnitude higher than the luminosity that is based on the narrow component. However, the \citet{netzer09} estimates of the bolometric luminosity of LINERs and Seyfert 2 galaxies, are based on observations of many thousands of sources, on previous estimates of the luminosity (e.g. \citealt{heckman04}), and on detailed photoionization calculations. The underlying assumption is that the lines are emitted from an NLR whose conditions, most importantly density, dust content, and covering factor of the central source, are similar in all sources, albeit with a large scatter. The narrow core LINERs lines in J132401.63+454620.6 are consistent with these assumptions but this is not the case for the broad line components. These components have velocities that significantly exceed those observed in Seyfert 2 galaxies. The great similarity in emission line ratios between the broad components, and narrower emission lines in many Seyfert galaxies, is not surprising since any gas excited by an AGN-type ionizing SED, with ionization parameter similar to the one in Seyfert galaxies NLR (about $\mathrm{10^{-2}}$) will produce such line ratios. In addition, the line luminosity depends linearly on the covering factor of the gas which can be an order of magnitude larger than the covering factor of the NLR in typical Seyfer galaxies, which is estimated to of order 5-10\% (e.g. \citealt{netzer13}). Given this we suggest that the broad, Seyfert-like components originate from fast moving gas which covers most of the central ionizing source. The calculations detailed below strengthen this suggestion. Our best estimate of the bolometric luminosity of the central source is hence derived from the LINER-like lines and is roughly $L_{bol}=10^{45}\,\mathrm{erg\,s^{-1}}$.

The S$\mathrm{\acute{e}}$rsic index of the galaxy is 4.3 in the $r$-band and 5.1 in $g$-band, therefore the galaxy is bulge-dominated, and we can estimate the BH mass using the stellar velocity dispersion. We use the M-$\sigma$ relation from \citet{kormendy13} and find a BH mass of $\mathrm{log\,M/M_{\odot}=7.67\pm0.3}$, where we assume a nominal uncertainty of a factor 2 since we measure the velocity dispersion of the entire galaxy. For solar metallicity, $L_{\mathrm{Edd}}=1.5\times10^{38}\mathrm{(M_{BH}/M_{\odot})\,erg\,s^{-1}}$ thus, from the LINER component, $L/L_{\mathrm{Edd}}= 0.14$. This Eddington ratio is consistent with Type II Seyferts and type-I QSOs but not with type II LINERs \citep{netzer09}. The accretion rate, assuming mass-to-radiation conversion efficiency $\eta_{acc}=0.1$, is $\dot{m}_{acc}=0.17\,$ \Msol/yr. 

\citet[hereafter PO16]{povic16} studied a sample of the 42 most luminous, highest SFR, LINERs in the local universe. They measure bolometric luminosities in the range $\mathrm{log}L_{bol}=43.95$ to $\mathrm{log}L_{bol}=45.48$. SDSS J132401.63+454620.6 shows bolometric luminosity within this range. To compare the star formation luminosity (LSF) to that of PO16, we follow their conversion of SFR to LSF: $\mathrm{LSF=SFR\times 10^{10}L_{\odot}}$, which is based on the Kroupa initial mass function. For our measured SFR of 25 \Msol/yr, we find $\mathrm{LSF=9.6 \times 10^{44}\,erg\,s^{-1}}$, which is higher than most of the values derived by PO16 for their sample, except for the most luminous LINERs. PO16 also bin their objects by AGN luminosity and measure the median LSF in each bin. For the bin that corresponds to the bolometric luminosity that we measure, their median LSF is about a factor 3 lower than what we measure. 

\citet[hereafter CA13]{cales13} studied the optical spectra of 38 type-I (broad emission lines) PSQs, 29 of which have morphological classification from the Hubble Space Telescope. They measure BH mass, accretion rate, and AGN luminosity for their sources. In a follow up work, \citet{cales15} compare their objects to post starburst galaxies and to QSOs, and find that PSQs with more luminous AGNs and more luminous post starburst stellar populations (compared to the observed old stellar population) are hosted by elliptical and early-type galaxies, some of which are clearly merger products. By comparing the spectra of PSQs to that of post starburst galaxies and QSOs, they argue that PSQs may be transitioning objects between typical post starburst galaxies and typical quasars. They further suggest that PSQs may be the descendants of ULIRGs. 

It is useful to compare SDSS J132401.63+454620.6 to the sources studied by CA13. The BH mass that we measure corresponds to the lowest values found by CA13, and the Eddington ratio corresponds to the highest values they find, both fit well within their BH mass-Eddington ratio correlation (see their figure 3). The morphology is consistent with an early type galaxy, and the starburst age we derive is lower than found by CA13, which may indicate that we observe a galaxy at an earlier transition stage than the galaxies in CA13. We measure bolometric luminosities using the BH mass and the Eddington ratio given in CA13 (see table 6 there). The distribution is log-normal with mean $\mathrm{log}L_{bol}\sim44.8$, the minimal and maximal values are $\mathrm{log}L_{bol}\sim44.2$ and $\mathrm{log}L_{bol}\sim45.4$ respectively. With its measured bolometric luminosity, and other properties, SDSS J132401.63+454620.6 would be classified as a PSQ by CA13, except that it is a type-II AGN while all objects in their sample are type-I AGN.

\subsubsection{Modelling the photoionized gas in SDSS J132401.63+454620.6}\label{s:photo}

Given the observed and reddening corrected line intensities, and the various line ratios, we can make simple estimates of the properties of the emission line gas. The simplified model presented
here is meant to test the main observed properties and is, by no means a complete photoionization model. In particular, we use a ``single cloud'' approach, where we calculate the properties of one cloud, with a given density, column density, and location with respect to the central ionizing source, as representing an entire emission line region. In many cases of interest, this is a poor replacement for a more complex model, that accounts for the geometrical distribution of clouds, a range in density, column density and location (see \citealt{netzer13} for a more complete description of multi-component photoionization models). The cloud is assumed to be dusty, with solar metallicity, and with an ISM-type depletion. The code used for the calculations is ION, most recently described in \citet{mor12}.

The properties of the central source of radiation follow standard assumptions about AGN SEDs. It is made of a combination of an optical-UV continuum emitted by a geometrically thin accretion disk and an additional X-ray power-law source extending to 50 keV with a photon index $\Gamma=1.9$. The normalisation of the UV (2500\AA) to X-ray (2 keV) flux is defined by $\alpha_{OX}$, which we chose to be 1.1. The total luminosity, based on the observations, and probably correct to within a factor $\sim 2$ (equation \ref{eq:1}), is assumed to be $10^{45}\,\mathrm{erg\,s^{-1}}$. For this continuum, the fraction of the ionizing luminosity, beyond the hydrogen Lyman continuum, is 70\%, and the number of ionizing photons is Q(Lyman)=$\mathrm{10^{54.7}\, sec^{-1}}$. For the LINER component we chose $n_{\mathrm{H}}=1000\,\mathrm{cm^{-3}}$ gas at a distance of $10^{21.8}\,\mathrm{cm}$ from the center. The covering fraction of this component, estimated from the reddening corrected luminosities of the Balmer lines, is 0.04. For the outflowing component we chose $n_{\mathrm{H}}=1000\,\mathrm{cm^{-3}}$, a distance of $10^{21.3}\,\mathrm{cm}$, and a covering fraction of 0.5. While the overall covering factor of the outflow is probably larger, given the red and blue wings of the $\mathrm{[OIII]}$ line, the model is based on the luminosities of the blue wings only and hence the chosen covering factor. The column density in both cases is large enough ($>10^{21.5}\,\mathrm{cm^{-2}}$) to result in complete absorption of the ionizing radiation. 

With these assumptions, the ionization parameter $\mathrm{U = Q(Lyman)} / 4\pi r^{2} n_{\mathrm{H}} c$, where $n_{\mathrm{H}}$ is the hydrogen number density and $c$ the speed of light, can be found for both components. For the LINER gas we find $\mathrm{U} = 3.8 \times 10^{-4}$ and for the outflowing gas $\mathrm{U} = 3.8 \times 10^{-3}$. This choice of parameters reproduces, to within a factor $\sim 2$, most of the observed properties, including the reddening corrected luminosities of the strongest lines in both components. Being a dusty gas, some of the line attenuation is the result of absorption and scattering inside the ionized gas and some require additional foreground dust either outside the line emission region or in neutral atomic gas at the back of the ionized cloud.

All the above properties are quite standard for photoionized AGN gas, except for the covering fraction of the outflowing gas, which is a factor of $\sim 5$ larger than the covering factor of the NLR in Seyfert galaxies. For the purpose of computing the gas and dust mass (see sections \ref{s:agn_quench} and \ref{s:dust_mol} below) we note that for a covering factor of unity and the chosen column density of $10^{21.5}\,\mathrm{cm^{-2}}$, the mass of the outflowing gas is about $1.8 \times 10^8\,\mathrm{M_{\odot}}$. Obviously, for a given column density and covering factor, the total mass scales with $r^2$ and a much smaller mass, with the same ionization parameter, can be achieved by scaling the gas density.

\section{Discussion}\label{s:disc}

Our analysis of the observed and derived properties of SDSS J132401.63+454620.6 suggests several unique features of this source, which we briefly summarize here: 
\begin{enumerate}
\item The galaxy has a narrow stellar age distribution (mass-weighted average age is 0.25 Gyr with a standard deviation of 0.11 Gyr) and no contribution from O and B-type stars. Thus, the galaxy is a typical E+A post starburst galaxy. 
\item The stellar population is resolved in the 2D ESI spectrum, and its half-light radius is 2 kpc. The star-light is reddened with a mean $\mathrm{E}(B-V)=\mathrm{0.3\,mag}$. 
\item The emission lines show two distinct velocity components, one is consistent with LINER-type emission (FWHM $= 320\,\mathrm{km\,s^{-1}}$) and the other with Seyfert-type emission (FWHM $= 1300\,\mathrm{km\,s^{-1}}$). The reddening towards the two components are different, corresponding to $\mathrm{E}(B-V)=\mathrm{0.68\,mag}$ and $\mathrm{E}(B-V)=\mathrm{1.82\,mag}$, respectively. Neither component is resolved in the 2D spectrum, with an upper limit of 1.4 kpc. 
\item We modelled the emission line gas by considering shock excitation, post-AGB ionization, and AGN ionization. We find that the former two are inconsistent with the observations, and conclude that both of the gas components are ionized by an AGN. Our photoionization modelling of the gas suggest a typical LINER component with a covering fraction of about 0.04, and a Seyfert-like outflowing component with a covering factor of $\sim0.5$. 
\end{enumerate}

The above results suggest that SDSS J132401.63+454620.6 is at a critical, short-lived, phase. The lack of ongoing star formation and the observed fast outflows suggest that AGN-driven winds quenched the recent starburst, and we are witnessing the final stages of the process, before the AGN expells the remaining gas in the galaxy and exhausts its gas. The following is an attempt to construct a model that explains our measurements and connects our results with the general properties of E+A galaxies.

\subsection{AGN feedback and SF quenching}

\subsubsection{Mass outflow rate}\label{s:agn_quench}

E+A galaxies are thought to be systems where a significant star formation episode has been quenched abruptly \citep{goto04, yang04, yang06, kaviraj07, cales13, cales15}. Some studies suggest an evolutionary path in which gas-rich major mergers result in ULIRGs, which then evolve to become post starburst galaxies (e.g., \citealt{kaviraj07}, \citealt{cales15}, \citealt{wild16}). The fate of such systems is to continue evolving into normal elliptical galaxies \citep{wild09, yang06, wild16}. The quenching of star formation in these systems was studied in detail by \citet{kaviraj07}. Their observations showed bimodal behaviour, and they argued that post starburst galaxies less massive than $\mathrm{10^{10}\,M_{\odot}}$ are quenched by feedback from supernovae, while more massive galaxies are quenched by AGN feedback. \citet{kaviraj07} noted the similarity between this threshold mass and the minimum mass above which AGN are significantly more abundant in nearby galaxies. However, so far there is no direct evidence for AGN feedback in E+A galaxies with masses exceeding $\mathrm{10^{10}\,M_{\odot}}$.

SDSS J132401.63+454620.6 seems to be peculiar among E+A galaxies in showing velocity dispersion in the Seyfert component which exceeds the escape velocity of the galaxy (section \ref{s:gas}). We interpret the broad emission lines as outflow of the ionized gas which is different from the outflows of cold ISM gas that were previously observed in post starburst galaxies (e.g., \citealt{tremonti07}). The stellar mass of the galaxy which is beyond the threshold suggested by \citet{kaviraj07}, and the direct evidence for AGN activity, are all consistent with AGN quenching in this system. To the best of our knowledge this is the first evidence for such quenching in an E+A galaxy.

To estimate the quenching mass outflow rate, we follow the analysis of \citet{soto12}. The outflow mass can be expressed as:
\begin{equation}\label{eq:2}
	{M_{out} = \mu m_{\mathrm{H}} V n_{e} f,}
\end{equation}
where $f$ is the filling factor of the gas, $n_{e}$ is the electron density, $V$ is the volume of the emitting region, $m_{\mathrm{H}}$ is the mass of the hydrogen atom, and $\mu$ the mass per hydrogen atom, which we take to be 1.4. The main uncertainty in this expression is the product $f\,n_{e}$, which is related to the emission line luminosities. For $\mathrm{H\alpha}$:
\begin{equation}\label{eq:3}
	{L_{\mathrm{H\alpha}} = \gamma n_{e}^{2} f V,}
\end{equation}
where $\gamma$ is the effective line emissivity which, for a recombination line in highly ionized gas, depends slightly on the electron temperature. For case B recombination and $T_{e} \sim 10^{4}\,\mathrm{K}$, $\gamma = 3.56 \times 10^{-25}\,\mathrm{erg\,cm^{3}\,s^{-1}}$ \citep{osterbrock06}. This is consistent with our photoionization calculations (section \ref{s:photo}). Using equations \ref{eq:2} and \ref{eq:3}, we get:
\begin{equation}\label{eq:4}
	{M_{out} = M_{0}\frac{L_{\mathrm{H\alpha}} }{n_{e}},}
\end{equation}
where $M_{0}$ is constant.

To estimate the mass outflow rate we define $t_{out}= r_{out}/v_{out}$, and take the velocity of the peak of the broad blueshifted emission relative to the peak of the narrow ($130\,\mathrm{km\,s^{-1}}$) added to the width of the broad blueshifted emission, $v_{out} = v_{peak, broad} + \sigma_{broad} = 700\,\mathrm{km\,s^{-1}}$. We choose this definition in order to be consistent with various outflow studies in ULIRGs.
Since the broad component is not spatially resolved in our 2D spectrum, we take the upper limit on the radius of the emitting gas to be determined by the seeing, i.e. $r_{out} = 1.4\,\mathrm{kpc}$. Since $\mathrm{\dot{M}_{out}}= M_{out} / t_{out}$, we get a lower limit on the mass outflow rate. The main uncertainty is the electron density. Using the broad $\mathrm{[SII]}$ line ratio (section \ref{s:gas}) we find $n_{e} = 100\,\mathrm{cm^{-3}}$. We use this estimate as the fiducial electron density of the emitting gas and the reddening corrected $\mathrm{H\alpha}$ luminosity listed in table \ref{t:meas}. This gives $\mathrm{M_{out} \approx 4 \times 10^{8}\, (100/n_{e})\, M_{\odot}}$ and,
\begin{equation}\label{eq:mdot}
 {\mathrm{\dot{M}_{out} \approx 300\, \Big(\mathrm{\frac{1 \,kpc}{r_{out}}} \Big) \, \Big(\frac{100\, cm^{-3}}{n_{e}} \Big) \, M_{\odot}/yr}.}
\end{equation}

We can now require an ionization parameter which results in the correct line ratio, U=$10^{-2.4}$ (see section \ref{s:photo}). We therefore introduce an additional constraint, on the product $n_{\mathrm{H}}\,r^{2}$ which is the one found from the photoionization model, i.e. $n_{\mathrm{H}} \, r^2 = 10^{45.6}\, \mathrm{cm^{-1}}$. Substituting in the mass outflow rate equation and expressing $r$ in units of kpc, we get: 
\begin{equation}\label{eq:mdot_final}
\mathrm{\dot{M}_{out}} \approx 83\, \Big(\mathrm{\frac{r}{1\,kpc}}\Big)\, \mathrm{M_{\odot}/yr}.
\end{equation}
Since r ranges between 1.4 kpc (spatial resolution limit) and about 50 pc (photoionization limit on the gas density, see previous discussion) we can constrain the mass outflow rate to be between about 4 and 120 \Msol/yr. The kinetic power of the outflow is therefore $6\times 10^{-4}\,L_{\mathrm{bol}}$ and $0.02\,L_{\mathrm{vol}}$ respectively. The outflow properties found for our source are in the general range found by \citet{fiore17} for the entire population.

A second estimate of the mass outflow rate can be obtained by using the [OIII]~$\lambda$5007\AA\, emission line \citep[see appendix A there]{cano12}, with similar assumptions about the gas properties. In this case:
\begin{equation}\label{eq:5}
	{\dot{M}_{out} = \frac{C L_{\mathrm{[OIII]} v_{out}}}{n_{e} 10^{\mathrm{[O/H]}} r_{out}},}
\end{equation}
where $C=\langle n_{e} \rangle^{2}/ \langle n_{e}^{2}\rangle$ is the condensation factor, which is related to the filling factor of the emitting gas. We estimate C=1, thus all ionized gas clouds have the same density. $10^{\mathrm{[O/H]}}$ is the oxygen abundance, which we measure to be close to solar. Following the same arguments, we derive a mass outflow rate which is about 30\% larger than the one derived from the broad $\mathrm{H\alpha}$ line.

ULIRGs, which are believed to be the result of gas-rich major mergers, show significant SFRs ($\mathrm{> 100\, M_{\odot}/yr}$) and significant mass outflow rates. \citet{soto12} studied a sample of 39 ULIRGs and argued that the high-velocity, ionized, emission-line gas is probably excited by shocks. They measured mass outflow rates that range between $\mathrm{0.6\, M_{\odot}/yr}$ and $\mathrm{80\, M_{\odot}/yr}$. Our measured mass outflow rate is within this range, but the SFR is lower than what is typically measured in ULIRGs (e.g., \citealt{rupke02}, \citealt{martin05}, \citealt{martin06}, \citealt{soto12b}, and references therein). 

\subsubsection{Quenching timescales}\label{s:dust_mol}

The evidence found here for a massive outflow is consistent with the idea that post starburst galaxies are the descendants of ULIRGs, and that for a galaxy mass that exceeds $\mathrm{10^{10}\,M_{\odot}}$, the quenching of star formation is dominated by AGN feedback \citep{kaviraj07}. In such a case, star formation is quenched by the AGN during the final ULIRG stages, and molecular gas and dust are expelled or destroyed by the AGN. Once the main starburst has ceased, and the dust reddening decreases, the merger remnant can morph into a post starburst galaxy. At this stage, there still exists enough gas to continuously fuel the BH accretion, and the AGN remains active until it exhausts its gas reservoirs. We therefore expect to find massive post starburst galaxies ($\mathrm{> 10^{10}\,M_{\odot}}$) with mass outflow rates that are consistent with those observed in massive ULIRGs, but with
reduced SFRs, since quenching is already taking place.

We also examine the timescales involved in removing the gas reservoir by star formation and by AGN-driven outflows. Our goal is to measure the gas mass in SDSS J132401.63+454620.6, and compare it to the gas mass typically observed in ULIRGs with similar properties. Using the measured constraint on the time since the starburst, we can evaluate the importance of AGN-driven outflows in removing gas from the ULIRG stage to the observed E+A stage.

\begin{figure*}
\includegraphics[width=0.462\textwidth]{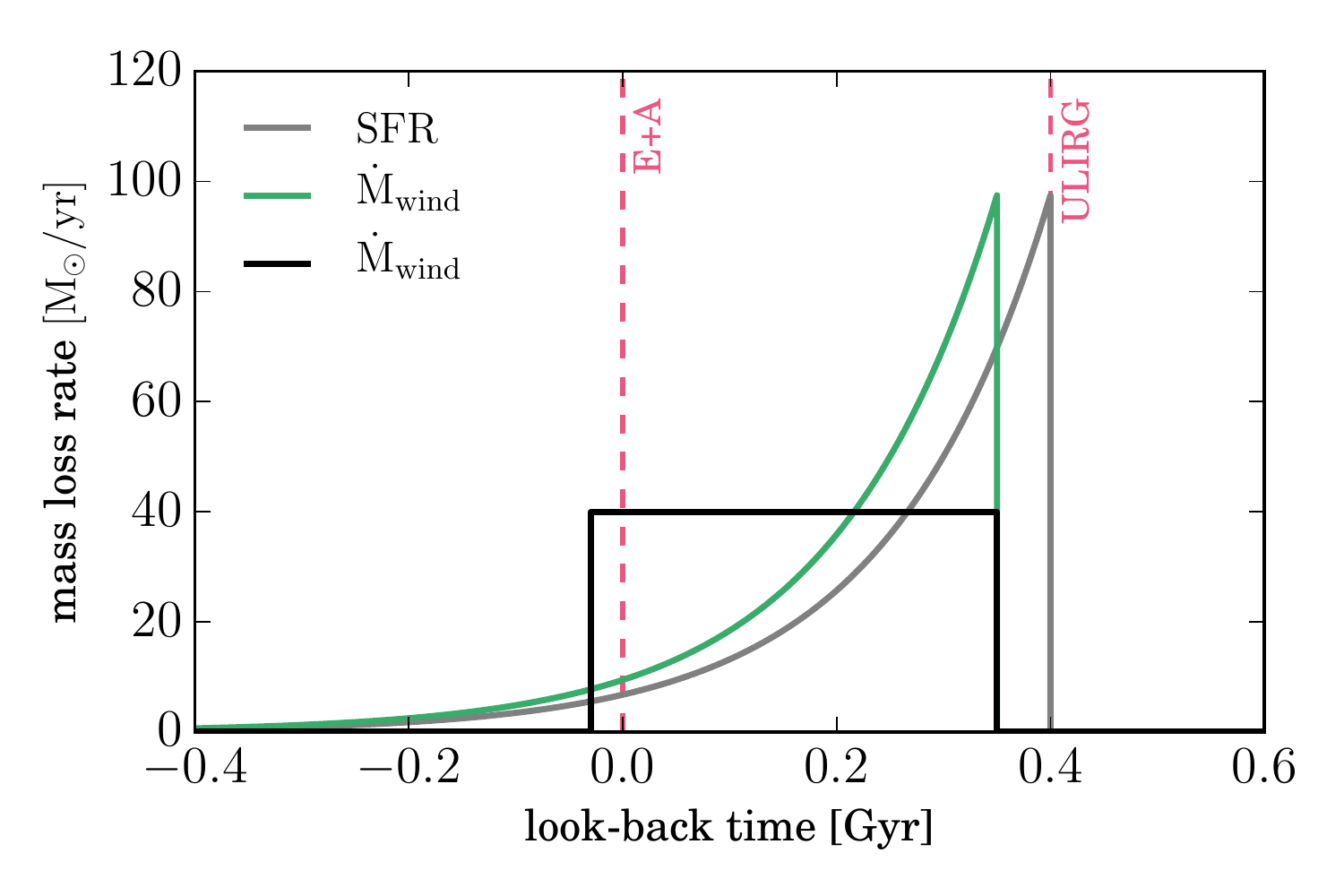}
\includegraphics[width=0.49\textwidth]{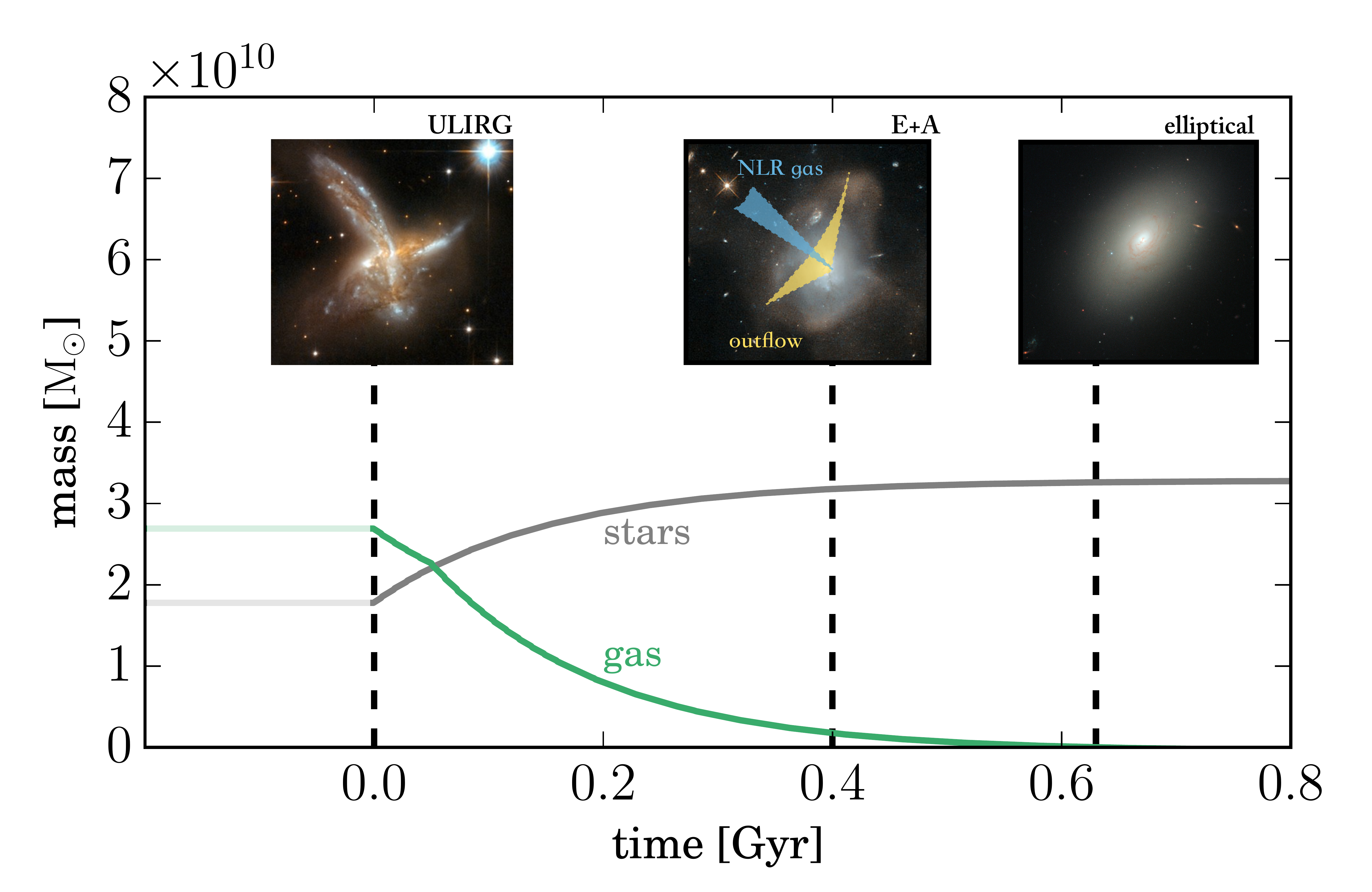}
\caption{\textbf{Left: }The assumed star formation history in SDSS J132401.63+454620.6 (black). Such history will remove roughly half of the gas mass in J132401.63+454620.6. The additional gas removal by AGN-driven outflows are either a constant mass outflow rate (grey) or mass outflow rate with shape that is similar to the SFH (green). \textbf{Right: } Our proposed model for J132401.63+454620.6. The gas (green) and stellar (grey) mass are shown as a function of time from the initial starburst at t=0 (ULIRG), through the current observed E+A phase at t=0.4 Gyr, to a red and dead elliptical at t$\sim$0.6 Gyr. We adopt the exponentially decaying model for the mass outflow rate.
}\label{f:feedback_cartoon}
\end{figure*}

We first model the SFH in SDSS J132401.63+454620.6 as $\mathrm{SFR(t)=SFR(t_b) \times \exp^{(t-t_b)/\tau}}$, where $\mathrm{t_b}$ is the time of the burst, $\mathrm{SFR(t_b)}$ is the SFR at the peak of the burst, and $\mathrm{\tau}$ is the decay time of the burst. The youngest star in our spectrum is 80 Myrs old, we therefore expect that the SFR will decay to less than 10\% of its initial value by the time we observe it (otherwise O and B-type stars would have been detectable in the observed SED). We construct 500 models with varying $\mathrm{t_d}$ and $\mathrm{\tau}$ using the MILES single stellar population synthesis models and obtain the parameters of the best fit to the observed spectrum. We find $\mathrm{t_d}$=400 Myrs and $\mathrm{\tau}$=150 Myrs. The peak star formation is $\mathrm{SFR(t_b)}\sim100\,\mathrm{M_{\odot}/yr}$, consistent with observed values in ULIRGs. Averaging over the last 100 Myrs, we get $\mathrm{SFR}\sim30\,\mathrm{M_{\odot}/yr}$, consistent with the value derived for our source in the JHU-MPA catalog. The total stellar mass resulting from the burst is $\mathrm{M_{*}=10^{10.13}\,M_{\odot}}$, which is 42\% of the total stellar mass in SDSS J132401.63+454620.6. This value is consistent with various estimates of stellar mass formed during bursts in post starburst galaxies (e.g., \citealt{kaviraj07}). We show in figure \ref{f:feedback_cartoon} the SFH of SDSS J132401.63+454620.6 according to this model (black curve).

Having the SFH, we can estimate the properties of the progenitor ULIRG of this system. \citet{genzel15} combine multiple observations of CO emission and Herschel far-IR dust measurements of over 500 star forming galaxies, and derive molecular gas masses using different methods. Their sample spans the redshift range 0--3, stellar masses of $\mathrm{M_{*}=10^{10-11.5}\, M_{\odot}}$, and the relative specific star formation rate (sSFR; see definition there) of $\mathrm{log(sSFR/sSFR(ms, z, M_{*}))}=-1.5 - 2$. A typical ULIRG is within this range. In order to estimate the gas mass using these scaling relations, we need to assume SFR, stellar mass, and redshift. We take the SFR at the peak of the burst $\mathrm{SFR(t_b)}\sim100\,\mathrm{M_{\odot}/yr}$, and redshift $z=0.15$ so that the look-back time between the progenitor and SDSS J132401.63+454620.6 ($z=0.125$) is 0.4 Gyr, which corresponds to the value of $\mathrm{t_d}$ that we measure. The total stellar mass formed in the burst is $\mathrm{M_{*}=10^{10.13}\,M_{\odot}}$, while the observed stellar mass of SDSS J132401.63+454620.6 is $\mathrm{M_{*}=10^{10.5}\,M_{\odot}}$. Thus the stellar mass of the progenitor is $\mathrm{M_{*}=10^{10.25}\,M_{\odot}}$. Using table 4 in \citet{genzel15}, and assuming that the total gas mass is 1.4 times the molecular gas mass, we obtain $\mathrm{M_{gas} \approx 10^{10.43}\, M_{\odot}}$. This value is consistent with gas mass observations in ULIRGs with similar properties (e.g., \citealt{solomon97}). 

The gas mass in SDSS J132401.63+454620.6 can be estimated from the observed reddening. We use the dust reddening that affects the starlight, $\mathrm{E}(B-V)=0.3\,\mathrm{mag}$, and assume that the dusty gas causing this reddening follows the stars. The typical size of this region is similar to the effective radius $r_{e}=2\,\mathrm{kpc}$ and, given the stellar distribution in a galaxy with a Sersic index of n=4.5, we expect a steep rise into the centre. We can test the consistency of this profile using the reddening towards the emission lines. The emission lines are not resolved in the 2D spectrum, with an upper limit radius of 1.4 kpc. The measured reddening towards the narrow and broad lines is $\mathrm{E}(B-V)=0.68\,\mathrm{mag}$ and $\mathrm{E}(B-V)=1.82\,\mathrm{mag}$ respectively (see section \ref{s:gas}). If we use the Sersic profile to estimate the radii at which the average dust reddening is $\mathrm{E}(B-V)=0.68\,\mathrm{mag}$ and $\mathrm{E}(B-V)=1.82\,\mathrm{mag}$, we find 1 and 0.45 kpc respectively. While we have no direct indication for such a smooth dust distribution, the numbers obtained here are consistent with the observations.

For solar metallicity and ISM-type depletion $\mathrm{n_{H} = \mathrm{E}(B-V) \times 5.6 \cdot 10^{21} \,cm^{-2}}$. Assuming a spherical gas distribution with a radius of 1 kpc and a mean $\mathrm{E}(B-V)=1.2\,\mathrm{mag}$ (between the two numbers derived above), we get a total gas mass of $\mathrm{M_{gas} \approx 10^{8.9}\, M_{\odot}}$. The uncertainty on this number is at least a factor of 3 and hence we adopt the value of $\mathrm{M_{gas} = 10^{9 \pm 0.5} \, M_{\odot}}$.

The simple SFH model we present here starts with a progenitor ULIRG with gas mass of $\mathrm{M(ULIRG)_{gas} = 10^{10.43}\, M_{\odot}}$, and ends with a post starburst galaxy with gas mass of $\mathrm{M(E+A)_{gas} = 10^{9.0}\, M_{\odot}}$. Therefore, one must remove a total gas mass of $\mathrm{M(t=0.4\, Gyr)_{gas} = 10^{10.41}\, M_{\odot}}$ in 0.4 Gyrs to evolve the assumed ULIRG into the observed post starburst galaxy, and then remove gas mass of $\mathrm{M(E+A)_{gas} = 10^{9.0}\, M_{\odot}}$ to evolve SDSS J132401.63+454620.6 to a red and dead galaxy. The total gas mass that is removed by star formation ($\mathrm{M_{*}=10^{10.13}\,M_{\odot}}$) is roughly half of the mass that must be removed to evolve the ULIRG with the properties assumed here into the observed E+A galaxy. We therefore suggest that the observed AGN-driven outflow removes the remaining gas mass. 

We consider two simplistic possibilities for the history of the outflow. We first assume that the AGN-driven outflow is constant throughout the starburst, and its value is the measured mass outflow rate from section \ref{s:agn_quench}. For example, assuming mass outflow rate of 40 \Msol/yr (in the middle of the range obtained earlier) will remove the remaining gas mass in 350 Myrs, and the AGN becomes active 50 Myrs after the peak of the starburst. Under this assumption, an additional time of 30 Myrs is necessary to remove the observed gas reservoir in SDSS J132401.63+454620.6. At this stage, the youngest stars one would observe in the spectrum are of age $\sim$100 Myrs. We show the constant mass outflow rate in the left panel of figure \ref{f:feedback_cartoon} (grey). 

A major concern of the above scenario is that the duration of the flow (350 Myr) is very long considering the expected lifetime of the AGN. We therefore tested a second, more realistic scenario that involves higher mass outflow rate in the past that decreases exponentially with time. There is a range of parameters in such functions and the only one we tried is shown in the left panel of figure \ref{f:feedback_cartoon} (green). In this case, the outflow starts 50 Myrs after the onset of the starburst. The initial mass outflow rate is $\mathrm{\dot{M}_{out}\sim 100 \, M_{\odot}/yr}$ and the timescale is 150 Myrs. The resulting shape is similar to the SFR function shown also in the same diagram. In this scenario, the wind removes the gas mass that is necessary to evolve the system to the present stage where the mass outflow rate is $\mathrm{\dot{M}_{out}=10\, M_{\odot}/yr}$, well inside the range obtained above. Assuming the AGN accretion rate is more-or-less constant during this time, we find that the BH mass increases by about a factor 2 during this part of the evolution. We show our proposed model in the right panel of figure \ref{f:feedback_cartoon}\footnote{Images credit: ESA/Hubble and NASA. Acknowledgements: Hubble Heritage Team (STScI / AURA)-ESA, A. Evans, J. Schmidt, R.M. Crockett, S. Kaviraj, J. Silk, M. Mutchler, and R. O'Connell}., where the gas and stellar mass are shown as a function of time from the initial starburst at t=0 (ULIRG), through the current observed E+A phase at t=0.4 Gyr, to a red and dead elliptical at t$\sim$0.6 Gyr. One may argue that even 150 Myr of continuous AGN activity is too long and the above mentioned time is broken into several shorter episodes. In this case, the mass outflow rate, as well as the SFR, must be scaled accordingly. We do not have enough information to distinguish between all these possibilities. 

Finally, we note that the gas mass estimate used here, which is based on a simple spherical distribution of the dusty gas, and the reddening of the broad emission lines, is of the same order of magnitude of the mass of the outflowing ionized gas estimated earlier (few $10^8$ \Msol for a distance of about 0.5 kpc and a column density of $10^{21.5}\, \mathrm{cm^{-2}}$, see section \ref{s:photo}). The outflowing gas can have a large enough column of neutral gas on the back side of the ionized material, $\approx 10^{22}\, \mathrm{cm^{-2}}$, to fully explain the observed broad line attenuation. In this case, the outflowing mass is similar to the value of $10^9$ \Msol\, obtained earlier, i.e., the outflow carries most of the gas in the galaxy. 

\subsection{Outflows in type II AGN and E+A galaxies}\label{s:type2s}

We find massive, AGN-driven outflow in the spectrum of SDSS J132401.63+454620.6, with a mass outflow rate that matches those observed in ULIRGs. While SDSS J132401.63+454620.6 is the first E+A galaxy in which such winds are observed and characterised, outflows are common and observed in many other sources. AGN feedback is invoked to explain the properties of local massive galaxies, intracluster gas, and intergalactic medium (e.g., \citealt{silk98, churazov05, bower06, croton06, hopkins06, schawinski07, schawinski09, mccarthy10, gaspari11}). It is also proposed as a mechanism to regulate the growth of stellar and BH masses \citep{silk98, fabian99, benson03, king03, granato04, dimatteo05, springel05, debuhr12}. AGN-driven winds span a large range of galaxy properties. They are observed in ULIRGs (e.g., \citealt{rupke05, rupke13, spoon13, veilleux13, zaurin13, sun14, burillo15, fiore17}), in type I and type II AGN (e.g., \citealt{heckman81, feldman82, heckman84, greene05, nesvadba06, moe09, rosario10, greene11, cano12, arav13, mullaney13, fiore17}), and even in quiescent red and dead ellipticals \citep{cheung16}. 

Since SDSS J132401.63+454620.6 is classified as a type II AGN, it is worth noting and comparing our results with studies of outflows in such AGN. \citet[hereafter MU13]{mullaney13} analysed a sample of roughly 24\,000 optically selected AGN from the SDSS and perform a decomposition of the [OIII]~$\lambda \lambda$4959,5007\AA\, emission lines into narrow and broad components, where broad components were regarded as indication of outflow. They measure $\mathrm{FWHM_{avg}}$, which is the flux-weighted FWHM of the [OIII]~$\lambda$5007\AA\, line and find that 17\% of the AGN show $\mathrm{FWHM_{avg} > 500\, km\,s^{-1}}$ and about 1\% show $\mathrm{FWHM_{avg} > 1000\, km\,s^{-1}}$. The fractions of the total flux coming from the narrow and broad components were measured by assuming \emph{the same} dust reddening, though we clearly measure different reddening for the two components. In SDSS J132401.63+454620.6 we measure the $\mathrm{FWHM_{avg}}$ of the [OIII]~$\lambda$5007\AA\, line using the fluxes of the narrow and broad components with \emph{no dust correction} to be $\mathrm{FWHM_{avg}= 830 \, km\,s^{-1}}$. This value is higher than 98\% of the AGN in their sample. Furthermore, MU13 find that type II AGN in their sample display a less prominent blue wing in the [OIII]~$\lambda$5007\AA\, compared to type I, thus 98\% is a lower limit. Therefore, SDSS J132401.63+454620.6 is rather extreme compared to the general type II population.

\citet[hereafter HA14]{harrison14} used integral field spectroscopy (IFU) to study 16 type II AGN, selected from the MU13 sample with $\mathrm{FWHM > 700\,km^{-1}}$ and in which the broad component contributes at least 30\% to the total flux of the [OIII]~$\lambda$5007\AA\, line. They derive AGN bolometric luminosities for their objects (in the range $\mathrm{L_{AGN}=10^{45}-10^{46}\,erg\,s^{-1}}$). The bolometric luminosity of SDSS J132401.63+454620.6 corresponds to the lowest values in this range. Their wavelength coverage includes only $\mathrm{H\beta}$ and [OIII]~$\lambda \lambda$4959,5007\AA. Therefore, they do not correct their [OIII]~$\lambda$5007\AA\, luminosities for dust extinction. Indeed, their [OIII]~$\lambda$5007\AA\, luminosities are generally one order of magnitude weaker than what we measure for SDSS J132401.63+454620.6. They find gas velocities of 600--1500$\,\mathrm{km\,s^{-1}}$ with observed spatial extents of 6--16 kpc. SDSS J132401.63+454620.6 shows comparable gas outflow velocity but, the spatial extent of this outflow is at most 1.4 kpc. Their wavelength coverage does not allow the characterisation of the stellar properties, specifically age, thus it is not clear how similar SDSS J132401.63+454620.6 is to the HA14 sample. Nevertheless, by examining the 1D SDSS spectra of the HA14 sample, we find that one galaxy (SDSS J121622.73+141753.0) exhibits an E+A spectrum, though less extreme. This objects shows clear signs of broad emission in other forbidden lines, such as $\mathrm{[NII]}$, $\mathrm{[OI]}$, and $\mathrm{[SII]}$, which can allow a more certain mass outflow rate measurement. 

While SDSS J132401.63+454620.6 has similar properties to some type II AGN, the study of AGN feedback in the context of galaxy quenching and evolution using post starburst E+A galaxies offers an advantage over other samples. Post starburst E+A galaxy spectra are dominated by stars with approximately a single age, and in many cases the stellar mass that was formed in the recent burst is a non negligible fraction of the entire stellar mass. This may allow the study of outflow evolution as a function of time, when time is measured through the age of the stars \citep{wild10}. Since these galaxies do not currently form stars, it is possible to disentangle feedback processes that are due to star formation from those due to AGN activity. 

\section{Conclusions}\label{s:concs}
We present new observations of SDSS J132401.63+454620.6, a post starburst galaxy at redshift $z=0.125$ which was discovered as an outlier by the anomaly detection algorithm of \citet{baron17} due to its strong Balmer absorption lines and broad forbidden and permitted emission lines. The new Keck/ESI spectrum presented here allows us to measure the stellar and emitting gas properties in the object. Our results can be summarised as follows: 

\begin{itemize}
\item We find a narrow stellar age distribution (mass-weighted average age is 0.25 Gyr with a standard deviation of 0.11 Gyr) and no evidence for ongoing star formation, confirming that SDSS J132401.63+454620.6 is a post starburst galaxy. A simple model fitting, which is by no means unique, suggests a starburst that started about 0.4 Gyr ago with an exponential decay time of 150 Myr. According to this model, 40\% of the observed stellar mass was formed during the burst, with a peak SFR of about $\mathrm{\sim 100\, M_{\odot}/yr}$, similar to SFRs observed in typical ULIRGs.

\item Fitting of the observed line profiles require both narrow ($\mathrm{FWHM \approx 320\, km\,s^{-1}}$) and broad ($\mathrm{FWHM \approx 1300\, km\,s^{-1}}$) components. Those components are seen in many emission lines: $\mathrm{H\alpha}$, $\mathrm{H\beta}$, $\mathrm{[OIII]}$, $\mathrm{[NII]}$, $\mathrm{[SII]}$, $\mathrm{[OII]}$, and $\mathrm{[OI]}$. The broad emission line ratios suggest a Seyfert-type emission and the narrow line ratios are LINER-like. In most of the lines we detect only one, blueshifted broad component. However, in both $\mathrm{[OIII]}$ and $\mathrm{H\alpha}$ we detect an additional broad redshifted component. We also find that the velocity dispersion observed in the broad component exceeds the escape velocity in the galaxy, which we interpret as an outflow. This is the first reported case of ionized gas outflows in an E+A galaxy.

\item We compare the measured emission line ratios and luminosities to predictions of shocked-gas models and find large inconsistencies. We also find that the LINER component cannot arise due to photoionisation by post-AGB stars, thus all emission lines are due to an AGN radiation field. The LINER component is among the most luminous LINER observed.

\item We find a BH mass of $\mathrm{log\,M/M_{\odot}=7.67}$, accreting at a few percent of the Eddington luminosity. These values agree with the properties observed in post starburst quasars by \citet[CA13]{cales13}. The BH mass corresponds to the lowest values observed in CA13, the Eddington ratio corresponds to their highest values and the starburst age is lower than those in the CA13 sample. This may indicate that we observe a galaxy at an earlier transition stage than the galaxies in CA13. A simple photoionization model can account for the emission line properties of both the LINER and the higher-ionization, outflowing gas components.

\item We use the $\mathrm{H\alpha}$ and the $\mathrm{[OIII]}$ broad emission lines, combined with constraints from the photoionization model, and find mass outflow rate ranging from 4 to 120 \Msol/yr. These values are consistent with the mass outflow rates observed in typical ULIRGs, which tightens the evolutionary connection between ULIRGs and post starburst galaxies. We find that the star formation alone cannot remove the necessary gas mass to evolve our assumed ULIRG into the observed post starburst galaxy, and suggest that the observed AGN-driven wind is partially responsible for the quenching of star formation. We compare two simplified models of AGN-driven winds and show that these, together with the SFR, can consistently remove the gas mass and evolve the assumed ULIRG into the post starburst galaxy observed today, within the time estimated from the stellar ages. Such a wind can remove the remaining gas mass and exhaust the AGN gas reservoir, and evolve SDSS J132401.63+454620.6 into an elliptical, red and dead, galaxy within 30-200 Myrs.

SDSS J132401.63+454620.6 is the first to show ionised gas outflows among E+A galaxies. Objects of this type show narrow age distribution, which allows one to study the time scales of the starburst, and examine the dust and molecular content as a function of stellar age. The significant AGN-driven feedback we found suggests that the time scale in which such galaxies show powerful winds is short, thus only a handful of similar objects are expected to be found, compared to the general E+A and ULIRG population. The relative fraction of such galaxies can provide constrains on the efficiency of AGN-driven outflows to migrate such galaxies to the red sequence. We are currently involved in a detailed analysis of similar objects and results will be reported in a forthcoming publication. 

\end{itemize}

\section*{Acknowledgments}
We thank the anonymous referee for useful comments.
We further thank A. Dey, O. Gottlieb, T. Heckman, B. Menard, K. Rowlands, B. Trakhtenbrot, J. Silk, A. Sternberg, and D. Wylezalek for useful discussions regarding this work and manuscript. 

The spectroscopic analysis was made using IPython \citep{perez07}. We also used the following Python packages: pyspeckit\footnote{www.pyspeckit.bitbucket.org}, and astropy\footnote{www.astropy.org/}.

This work made use of SDSS-III\footnote{www.sdss3.org} data. Funding for SDSS-III has been provided by the Alfred P. Sloan Foundation, the Participating Institutions, the National Science Foundation, and the U.S. Department of Energy Office of Science. SDSS-III is managed by the Astrophysical Research Consortium for the Participating Institutions of the SDSS-III Collaboration including the University of Arizona, the Brazilian Participation Group, Brookhaven National Laboratory, Carnegie Mellon University, University of Florida, the French Participation Group, the German Participation Group, Harvard University, the Instituto de Astrofisica de Canarias, the Michigan State/Notre Dame/JINA Participation Group, Johns Hopkins University, Lawrence Berkeley National Laboratory, Max Planck Institute for Astrophysics, Max Planck Institute for Extraterrestrial Physics, New Mexico State University, New York University, Ohio State University, Pennsylvania State University, University of Portsmouth, Princeton University, the Spanish Participation Group, University of Tokyo, University of Utah, Vanderbilt University, University of Virginia, University of Washington, and Yale University.

\bibliographystyle{mn2e}
\bibliography{ref}

\end{document}